\documentclass[twocolumn,showpacs,floatfix,pre,superscriptaddress]{revtex4-1}%

\usepackage{graphicx} % Include figure files
\usepackage{dcolumn} % Align table columns on decimal point
\usepackage{epsf} % bold math
\usepackage{amsmath}
\usepackage{bm} % bold math
\usepackage{setspace}
\usepackage{xcolor}
\usepackage[switch]{lineno} 
\usepackage{cmap}
\usepackage{lmodern}
\usepackage[english]{babel}

\begin{document}

%%%%%%%%%%%%%%%%%%%%%%%%%%%%%%%%%%%%%%%
%%%%%%%%%%%%%%%%%%%%%%%%%%%%%%%%%%%%%%%

\title{Crystalline structures of particles interacting through the harmonic-repulsive pair potential}

\author{V.A.~Levashov}
\affiliation{Technological Design Institute of Scientific Instrument Engineering,
Novosibirsk, 630058, Russia}

%%%%%%%%%%%%%%%%%%%%%%%%%%%%%%%%%%%%%%%%%%%%%%%
%%%%%%%%%%%%%%%%%%%%%%%%%%%%%%%%%%%%%%%%%%%%%%%

\begin{abstract}
The behavior of identical particles 
interacting through the harmonic-repulsive pair potential 
has been studied in 3D 
using molecular dynamics simulations 
at a number of different densities.
We found that at many densities, 
as the temperature of the systems decreases,
the particles crystallize into complex structures 
whose formation have not been anticipated
in previous studies on the harmonic-repulsive pair potential.
In particular, at certain densities crystallization into the structure 
$Ia\bar{3}d$ (space group $\#230$) with 16 particles in the unit cell occupying 
Wyckoff special positions (16b) was observed.  
This crystal structure has not been observed previously in experiments or in computer simulations of single
component atomic or soft matter systems.
At another density we observed a liquid which is rather stable against crystallization. 
Yet, we observed crystallization of this liquid into the monoclinic $C2/c$ (space group $\#15$) structure  
with 32 particles in the unit cell occupying four different non-special Wyckoff (8f) sites.
In this structure particles located at different Wyckoff sites have different energies.
From the perspective of the local atomic environment, the organization of particles in this structure resembles the structure of some
columnar quasicrystals.
At a different value of the density we did not observe crystallization at all
despite rather long molecular dynamics runs.
At two other densities we observed the formation 
of the $\beta Sn$ distorted diamond structures 
instead of the expected diamond structure.
Possibly, we also observed the formation of the $R\bar{3}c$ hexagonal lattice with 24 particles 
per unit cell occupying non-equivalent positions.
\end{abstract}

\today

\maketitle

%%%%%%%%%%%%%%%%%%%%%%%%%%%%%%%%%%%%%%%%%%
%%%%%%%%%%%%%%%%%%%%%%%%%%%%%%%%%%%%%%%%%%

\section{Introduction}

If spherically symmetric pair potentials are used to model the behavior of the atomic systems then 
these potentials usually have steep repulsion at short distances and they diverge at zero 
separation between the model particles. 
The most common example is the Lennard-Jones potential often used to model the properties of gases, liquids, 
and solids of inert atoms. If crystal structures arise in the simulations with such 
hard-core pair potentials then they are usually Face Centered Cubic (FCC) or Body Centered Cubic (BCC) lattices.
\cite{HansenJP20061,Frenkel20021,Tadmor2011,VerletJP1969,Hoover19711,Frenkel19961}

Over the last thirty years, significant attention has been paid to 
the modeling of soft matter systems 
\cite{Kleman20031,Lang20001,Louis20001,Likos20011,Likos20012,Likos20021,Likos20061,
Malescio20071,Fomin20081,Malescio20081,Saija20091,Frenkel20091,Prestipino20091,
Malescio20111,LuZY20111,Jorjadze20131,Knobloch2013,Lifshitz2014,Mohanty20141,Xu20141,Xu20151}.
Some soft matter systems consist of star polymers or dendrimers, 
or micelles, or microgels in solutions. These mesoscopic particles have approximately 
spherical shape and they can be modeled with spherically symmetric 
pair potentials \cite{Likos20021,Likos20061,Mohanty20141,Malescio20071,LuZY20111,Knobloch2013,Lifshitz2014,Xu20141}.

The interactions between the particles in the soft matter systems are quite 
different from the interactions between atoms. 
The corresponding modeling pair potentials may have only repulsive 
parts (no attraction between the particles) which usually are much softer than the repulsive part met in the atomic interactions 
\cite{Kleman20031,Lang20001,Louis20001,Likos20011,Likos20021,Likos20061,
Malescio20071,Malescio20081,Fomin20081,Saija20091,Likos20011,Malescio20111,
Knobloch2013,Jorjadze20131,Lifshitz2014,Mohanty20141}.
Moreover, complete overlap of mesoscopic particles is sometimes 
possible and correspondingly some 
modeling potentials have finite value at zero separation distance between the particles 
\cite{Kleman20031,Likos20011,Likos20021,Likos20061,Malescio20071,Mohanty20141,Knobloch2013,Lifshitz2014}.
At present, it is well known that the systems of particles interacting through soft potentials 
can form rather unusual structures in comparison to the structures formed by particles 
interacting through simple spherically symmetric atomic ``hard-core" pair potentials 
\cite{Kleman20031,Likos20021,Likos20061,Malescio20071,Mohanty20141,Knobloch2013,Lifshitz2014}. 

Discovery of quasicrystals lead, in particular, to the systematic studies of the single 
component systems consisting of the particles interacting through spherically-symmetric 
pair potentials whose shape is more complex than the shape of the simple ``traditional" pair potentials 
\cite{Schechtman19841,Steinhardt19841,Steurer20121,Janot20121,Olami19901,
Dzugutov19931,Engel20071,Engel20151,Ryltsev20151,Ryltsev20171,Damascento20171}.
The goal of the related studies is often to clarify the relationship between 
the shape of the potential and the structural/dynamic properties of the systems of particles interacting 
through such potentials.

Two other directions that concern the studies of the ``unusual'' 
interaction pair potentials are related to the general investigations of the ground states of 
pair potentials whose shape is constrained in some sense and the potentials 
that lead to non-crystalline ground states 
\cite{Radin1987,Torquato2015,Batten2009,
Stillinger20111,Stillinger20131}.

Finally, we mention yet another research direction which discusses 
the ``unusual" crystal structures, i.e., the direction that concerns 
the intent to design particles interacting through such potentials 
that would lead to the desirable properties
of the materials \cite{Stillinger20071,Cohn2009,
Torquato2009,Glotzer2013,Jaeger2015}.

In this article we report about several unexpected observations that have been made 
in our investigations of systems of identical particles interacting through the repulsive harmonic pair potential:
\begin{eqnarray}
\phi(r_{ij}) = \epsilon\left(1-\frac{r}{\sigma}\right)^2\theta\left(1-\frac{r}{\sigma}\right),
\label{eq:potential-1}
\end{eqnarray}
where $\theta(x)$ is the Heaviside step function, while $\epsilon$ and $\sigma$ 
determine the energy and length scales of the potential.

Originally we became interested in this potential because we wanted to 
address the generality of a particular observation described in Ref.\cite{Levashov20162,Levashov20162c}.

Behavior of particles interacting through potential (\ref{eq:potential-1}) has 
been investigated before in several different contexts in a number of previous publications \cite{Mohanty20141,Nagel20091,Berthier2010,
Zamponi2011,LuZY20111,Xu20141,Xu20151}. 
In particular, in Ref.\cite{LuZY20111} the phase diagram of particles 
interacting through potential (\ref{eq:potential-1}) has been constructed.

In our MD simulations we observed the formation of crystalline structures 
which are noticeably more complex than the structures considered in Ref.\cite{LuZY20111}. 
See also Ref.\cite{Xu20141}. 

The method used in Ref.\cite{LuZY20111} to construct the phase diagram consists of three steps. 
In the first step, the set of the possible crystal structures is assumed. 
In the second step, the set of the considered crystal structures is narrowed through 
the check of their stabilities at the density of interest (at low temperatures) 
using the method of dissipative particle dynamics (after all,
it is a particular method of molecular dynamics).
In the third step, the phase diagram of the system is constructed 
through calculations and comparisons of the free energies of 
the considered crystal structures.

The analysis implemented in Ref.\cite{LuZY20111} previously was also used in Ref.\cite{Frenkel20091} 
to construct the phase diagram of particles interacting through the repulsive Hertzian potential. 
The zero-temperature phase diagram of particles interacting through the Hertzian potential 
has been also studied in Ref.\cite{Prestipino20091}. 
In Ref.\cite{Prestipino20091} a significantly larger set of the crystal structures 
has been considered as possible in comparison to the set of structures studied in Ref.\cite{Frenkel20091}. 
The results presented in Ref.\cite{Prestipino20091} suggest that the actual phase 
diagram of particles interacting through the Hertzian pair potential 
is more complex than the phase diagram obtained in Ref.\cite{Frenkel20091}. 
Given the situation with Ref.\cite{Frenkel20091,Prestipino20091}, our observation 
of the behavior which is more complex than the one described is Ref.\cite{LuZY20111} is not really surprising.
However, some of the features that we observed in the behavior of systems of particles 
interacting through the harmonic-repulsive
potential have not been anticipated for such a simple interaction potential. 
In particular, we observed the formation of complex crystalline states with particles
occupying non-equivalent sites. One of these structures resembles, from a local perspective,
the structures of columnar quasicrystals. 
We also observed, at some densities, liquid states which are unexpectedly 
stable against crystallization on cooling.

In our approach, we acted in a direct way that does not rely on any initial assumptions.
 
We used LAMMPS molecular dynamics program \cite{Plimpton1995,lammps} and traditional 
MD simulations to produce liquid states through melting of the initial FCC or BCC crystals. 
The initial structure is of no importance after the melting. 
Then we cooled the liquid produced in this way. Initially we 
performed simulations at the density $\rho_o \sigma^3 =3.352$ and
found that at some temperature the liquid crystallizes into the structure whose pair density 
function, after cooling to nearly zero temperature, is shown in Fig.\ref{fig:pdfs-01}(f). 
We note that on further cooling we did not observe any signature 
of a transition into a different crystal structure. 
We also did not observe any transition of the thus obtained 
crystal structure into a different structure on heating until melting. 
Our analysis of the obtained crystal structure is presented in section (\ref{sec:LLL-lattice-1}). 
The point that we would like make here is that this crystal structure is rather unusual and 
it has not been observed previously in experiments or in computer simulations of single 
particle systems \cite{according,Luzzati1968,Kutsumizu2012,Tschierske2013}. 
Then we also realized that our results are in disagreement with the results presented in Ref.\cite{LuZY20111}. 
This situation motivated us to investigate what happens at other densities. 
On the basis of our investigations, we can not claim that we calculated the phase diagram of the system. 
However, we investigated a rather large set of densities and found several surprising results. 
Our major findings are summarized in the abstract and in the concluding section of this paper.

The article is organized as follows.
In section (\ref{sec:MD-details}) we describe the details 
of the MD simulations and the methodology of the structure analysis.
In section (\ref{sec:results}) the results of the MD simulations and 
the structure analysis at different densities are presented.
In section (\ref{sec:groundstate}) we analyze the relative stability 
of the observed and some other crystal structures from the 
perspective of the ground state potential energy.
In section \ref{sec:NPT-simulations} we briefly discuss 
the dependence of the Gibbs free energy on pressure at zero temperature.
We conclude in section (\ref{sec:conclusion}).

\section{Details of MD simulations and Data Analysis}\label{sec:MD-details}

We performed molecular dynamics simulations (MD) using 
the LAMMPS program \cite{Plimpton1995,lammps}. 
The simulations were performed using the Lennard-Jones (LJ) units \cite{lammps-lj-units}. 
This choice of units determines the energy scale: if the value of the harmonic repulsive 
potential (\ref{eq:potential-1}) at zero separation is equal to one, i.e., $\epsilon=1$, 
then $\epsilon$ corresponds to the depth of the minimum of the LJ-potential in the LJ-units. 
If the length scale of the harmonic-repulsive potential is equal to one, i.e., $\sigma =1$, then 
this length corresponds to the particles' diameter associated with the LJ-potential. 

In the following, all results will be presented in the LJ-units in 
accordance with the LAMMPS conventions, i.e., the temperature, $T$, 
and the Potential Energy per Particle (PEpP) will be measured in 
the units of $\epsilon$. The time, $t$, will be given in the units 
of $\tau = \sqrt{\sigma^2 m/\epsilon}$ \cite{lammps-lj-units}.
 
The magnitude of the time step was determined in the constant energy runs (NVE-ensemble) 
so that the total energy of the system is conserved with high precision (essentially 
no variation in the sixth digit of the value of the total energy per particle). 
The data were acquired in the constant temperature runs (NVT-ensemble) with Nos\'{e}-Hoover 
thermostat. The value of the time step varied in the interval between $\delta t \sim 0.001\tau$ for 
high temperatures $T \sim 0.015$ and  $\delta t \sim 0.1\tau$ for nearly zero temperature $T \sim 0.000025$. 
The value of the damping parameter associated with the Nos\'{e}-Hoover thermostat was chosen to be equal to 100 time steps.

Most of our simulations were performed on systems containing 13500 particles in a cubic simulation box. 
Sometimes we used 16000 and 18522 particles. At the density $\rho_o \sigma^3 =3.352$ we also made 
simulations on a system containing more than 100000 particles. 
To test the guessed crystal structures with a non-cubic symmetry, 
we created the guessed structure in the corresponding 
non-cubic geometry of the simulation box 
and run the simulation in the non-cubic geometry of the box. 
Periodic boundary conditions were always assumed.
In our simulations we did not notice size effects. 
Correspondingly, in the following discussions, 
we usually will not mention the sizes of the systems 
on which the data were obtained.

\begin{center}
\begin{table*}
\begin{tabular}{| c | c | c | c | c | c | c | c | c |} \hline
1                            & 2                & 3                & 4               & 5     & 6                & 7             & 8                  & 9        \\ \hline
                & $T_c$        & $U_L(T_c)$  & $U_C(T_c)$   &   & $U_C(T=0)$       & $U_{CI}(T=0)$    & $T_m$         &  \\
$\rho_o \sigma^3$                   & $\cdot 10^3$ & $\cdot 10^2$& $\cdot 10^2$ & Crystal  & $\cdot 10^2$       & $\cdot 10^2$    & $\cdot 10^3$ & Agreement \\ \hline
$1.68$    & $7.50$ & $4.35$  & $3.45$ & $FCC$ & $2.12$ & $1.87$ & $15.00$ & Yes      \\ \hline
$1.84$    & $8.00$ & $7.15$  & $5.90$ & $FCC$ & $4.92$ & $4.23$ & $15.33$ & Yes      \\ \hline
$2.1112$  & $7.50$ & $11.20$  & $10.35$ & $BCC$ & $9.12$ &$9.03$ & $14.50$ & Yes      \\ \hline
$2.4000$  & $6.48$ & $16.82$ & $15.87$& $BCC$ & $14.81$ & $14.74$ & $10.00$ & Yes      \\ \hline
$2.9040$  & $2.50$ & $26.62$ & $26.42$& $C2/c$& $26.01$ & $ 25.95 $& $6.00$ & No   \\ \hline
$3.25$    & $3.75$ & $33.84$ & $32.95$& $Ia\bar{3}d$ & $32.34$ & $32.21$ & $8.88$  & No       \\ \hline
$3.3520$  & $3.75$ & $35.84$ & $34.76$& $Ia\bar{3}d$ & $34.29$ & $33.99$ & $10.00$ & No       \\ \hline
$3.9000$  & $5.38$ & $46.89$ & $45.63$& $Ia\bar{3}d$ & $44.62$ & $44.42$ & $12.50$ & No       \\ \hline
$4.1600$  & $5.00$ & $51.93$  & $50.99$& $Ia\bar{3}d$ & $50.14$& $49.93$ & $10.38$ & No       \\ \hline
$4.4000$  & $N/A$  & $N/A$    & $N/A$  & $N/A$ & $N/A$    & $N/A$    & $N/A$   & ?        \\ \hline
$4.5000$  & $N/A$  & $N/A$    & $N/A$  & $N/A$ & $N/A$    & $N/A$    & $N/A$   & ?        \\ \hline
$5.0000$  & $4.04$ & $68.87$ & $67.88$ & $I4_1/amd$&$67.17$& $67.09$ & $9.38$ & No     \\ \hline
$5.2400$  & $3.25$ & $73.46$ & $72.72$& $I4_1/amd\;(?)$&$72.19$& $67.09$ & $9.00$ & No     \\ \hline
$6.0880$  & $3.50$ & $90.57$ & $90.05$& $P6_{3}/mmc\;(?)$&$89.46$& $89.49$ & $7.13$& N/A    \\ \hline
$7.0000$  & $5.18$ & $109.73$  & $108.00$ & $P6_{3}/mmc$&$107.06$&$106.91$  & $11.25$& N/A   \\ \hline
$7.8000$  & $4.25$ & $125.71$ & $124.67$& $P6_{3}/mmc$&$123.95$ &$123.75$  & $10.00$     & N/A \\ \hline
$8.8000$  & $5.00$ & $146.19$ & $145.10$& $R\bar{3}c$ &$144.20$&$$144.03$$& $8.15$& N/A       \\ \hline
$9.6000$  & $6.25$ & $162.76$ & $161.67$& $BCC$&$160.51$ & $160.47$& $9.88$ & N/A         \\ \hline
\end{tabular}
\caption{
The 1st column shows the densities at which the NVT simulations were performed. 
The 2nd column shows the temperatures at which we observed crystallization on cooling. 
These, of course, are not the true ``crystallization" or ``melting" temperatures.  
In particular, these ``crystallization" temperatures depend on the cooling rate. 
Thus, these are simply those temperatures at which we observed crystallization.  
The 3rd column shows the approximate values of the potential energy per particle 
in the liquid just before the crystallization. 
The 4th column shows the approximate values of the potential energy per particle 
immediately after the abrupt stage of crystallization, 
but before the slow relaxation that follows. 
We note that these are the values obtained in one particular run. 
These values can be noticeably different in different runs. 
The 5th column shows the crystal structures 
that were guessed from the structures produced 
in MD simulations after crystallization, further relaxation, 
and cooling to nearly zero temperature (see the previous 
section on how the simulations were performed).
The 6th column shows the approximate values of the potential 
energy per particle of the crystal structures obtained 
from the liquid state after cooling to zero temperature.
We note that these are the values obtained in one particular run. 
These values can be noticeably different in different runs.
The 7th column shows the values of the potential energy 
per particle of the guessed crystal structures without 
defects at zero temperature. 
The 8th column shows the melting temperatures of the guessed crystal structures on heating. 
The 9th column shows if there is the agreement with 
the results of Ref.\cite{LuZY20111} at this density. 
``N/A" in the 10th column stands for ``Not Applicable" since in Ref.\cite{LuZY20111} 
these values of the densities have not been studied. $FCC$ and $BCC$ stand for 
the Face Centered Cubic and Body Centered Cubic lattices correspondingly.
The notation $C2/c$ stands for the monoclinic structure with 32 particles in the unit cell occupying four
different non-special Wyckoff (8f) sites.
The $Ia\bar{3}d$ (space group $\#230$) stands for the cubic 
lattice with 16 particles in the unit cell 
occupying the (16b) Wyckoff special positions.
$I4_1/amd$ (space group $\#141$)  stands for the tetragonal lattice with 4 particles 
in the unit cell occupying the (4b) Wyckoff special positions. 
This $I4_1/amd$ crystal is also the $A5$ and the $\beta Sn$ structure. 
The $P6_3/mmc$ (space group $\#194$) stands for the hexagonal 
lattice with 2 particles in the unit cell occupying the (2c) Wyckoff special positions. 
The $R\bar{3}c$ (space group $\#167$) stands for the hexagonal 
lattice with 24 particles per unit
cell occupying the (6a) and (18e) Wyckoff special positions.
}
\label{table:MD-summary-1}
\end{table*}  
\end{center} 

\subsection{Data collection procedure}

At all densities we followed the same procedure for data collection.

1) At first, we generated a system at the required density as FCC lattice. 
It does not matter if this FCC lattice is stable or not at this density as on the next step the system was 
heated above its melting temperature. The fact that the system is in a liquid state was monitored 
using the Pair Density Function (PDF) and the dependence of the Mean Square Displacement (MSD) of the particles on time. 
The dependence of the system's potential energy per particle (PEpP) 
on time was used to monitor weather the system has reached the equilibrium state. 
The PDFs of the liquids 
do not have the sharp peaks that correspond to the lattice spacings and 
the diffusion rate is very significant in comparison to a crystal state 
in which the diffusion process is nearly absent. 

2) Then the temperature of the liquid was reduced to some lower value. 
Sometimes we used an abrupt decrease in temperature (an instant drop) 
and sometimes we used some cooling rate. 
For our purposes, the way in which the temperature is reduced 
is of no significance if the liquid at the reduced temperature remains
a liquid with a high diffusion rate. 
In this case the system reaches its equilibrium state relatively 
quickly and the cooling history does not influence the properties 
of the equilibrium state after some relatively short time. 
We monitored that the equilibrium is reached using the dependence of the PEpP on time. 
We also monitored that the diffusion rate remains significant 
and that there do not occur noticeable changes in the PDF with time.

3) When step 2) is repeated several times relaxation to the equilibrium becomes noticeably slower. 
The diffusion process also slows down. 
In addition, the lineshape of the PDF starts to exhibit more features 
that reflect the development of some structural ordering. 
All these changes are well known from the simulations of liquids. 
The point that we would like to make here is that in our simulations crystallization 
from the liquid state usually happens when some slowness in the relaxation develops. 
Since our goal was to produce crystalline states we performed longer simulations of 
the liquid states at temperatures where the slowness in the dynamics is already present. 
This approach usually allowed us to observe crystallization.

4) If we observed crystallization at some temperature then the system was allowed 
to crystallize and relax for a significant amount of time at this ``crystallization" temperature. 
The relaxation of the system was monitored using the dependence of the PEpP on time. 
The crystalline states formed in this way can not relax completely and defects in 
the crystal structures always remain. However, after some time further relaxation becomes very slow. 
The PDFs of the nearly relaxed states can be calculated and they show clear qualitative 
differences with the PDFs of the liquid state before the crystallization 
(this is a well-known fact that we mention in order to describe how our simulations were performed). 
It was also observed that the diffusion process in the crystalline states is nearly absent.

5) At this stage the obtained crystalline states were cooled to nearly zero temperature using some finite cooling rate. 
This cooling rate was usually quite slow in order to eliminate as many defects as possible. 
Sometimes, in order to eliminate the defects, we performed longer runs at some fixed 
temperatures lower than the crystallization temperature.  
It was found that this approach sometimes indeed helps to produce 
sharper peaks in the PDFs and correspondingly more pronounced crystal structures.
See Ref.\cite{heat-cryst-1} for an additional comment.

6) After a crystal structure at the very low temperature was obtained we tried 
to determine what this crystal structure is.
For this, we used visual analysis of the crystal structure and the PDF (see section \ref{sec:visual}). 
After we guessed the structure we optimized its parameters in order to minimize its potential energy. 
Then we created the guessed and optimized structure as an input structure file 
for the LAMMPS program. Of course, this guessed structure does not have any defects. 
Then we run LAMMPS MD simulations on the guessed and optimized structure at low temperatures. 
If the guessed structure was stable in the MD simulations then we assumed that our guess might be correct.

%%%%%%%%%%%%%%%%%%%%%%%%%%%%%%%%%%%%%%
% Begin Figure
%%%%%%%%%%%%%%%%%%%%%%%%%%%%%%%%%%%%%%
\begin{figure*}
\begin{center}
\includegraphics[angle=0,width=7.0in]{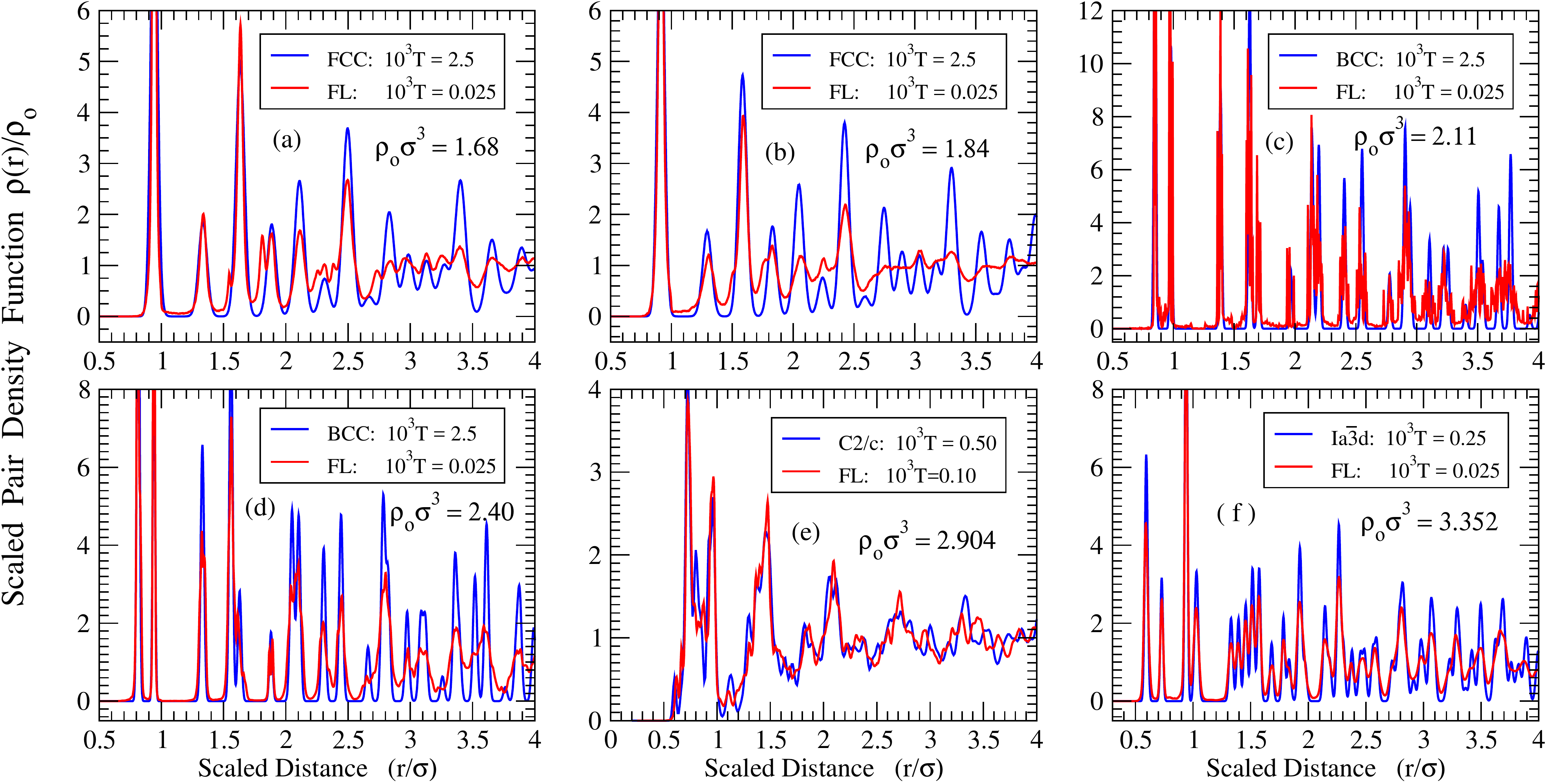}
\caption{The PDFs, $\rho(r)$, of the crystal structures obtained through crystallization from 
the liquid states and the PDFs of the corresponding guessed structures at the particles' 
densities shown in the panels. The ``FL" notation stands for ``From Liquid." 
Note that there is interaction between the central particle and all other particles which are closer to it than $(r/\sigma)=1$.
}\label{fig:pdfs-01}
\end{center}
\end{figure*}
%%%%%%%%%%%%%%%%%%%%%%%%%%%%%%%%%%%%%%
% End Figure
%%%%%%%%%%%%%%%%%%%%%%%%%%%%%%%%%%%%% 

7) On this step we heated the guessed crystal structure at some heating rate until its melts. 
Thus observed ``melting" temperature was usually significantly higher than the ``crystallization" 
temperature at the corresponding densities. 
This heating procedure provides another test for the correctness of the crystal structure guess. 
At all densities, except $\rho_o\sigma^3 = 8.8$, we did not observe transformations of the guessed 
structures into some other structures before melting.

8) We produced the crystal structures from the liquid states in the NVT simulations. 
In order to address further the stabilities of the obtained crystal structures, we also performed 
NPT simulations on the crystal structures obtained from the liquids. 
In these NPT simulations we varied the pressure at some selected constant temperatures. 
The stabilities of the crystal lattices were monitored through the dependencies of 
their potential energies on pressure. The abrupt changes in the derivatives of 
these dependencies were considered as indications of the lattice instabilities. 
The pressure ranges of the lattice stabilities will be discussed in section \ref{sec:NPT-simulations}.
This approach, of course, does not establish the pressure-temperature phase diagram. 
However, it does provide an additional insight into the ranges of the lattice stabilities.

\subsection{Visual analysis of the structures}\label{sec:visual}
   
We performed the visual analysis of the structures 
in two different ways.

One way was to select an atom and consider the geometry of its neighbor environment. 
Usually we considered several randomly chosen atoms from different regions of the simulation box. 
We also considered the results from the different runs. 
In almost all structures that we analyzed we found that the environments of all atoms are similar if the defects are ignored.

In a different approach we extracted from the whole simulation box some region and tried to guess 
the crystal structure from the structure of this region. Usually we considered several extracted regions of different sizes. 
This helped us guess the structure and also served as a check for the correctness of our guess.

%%%%%%%%%%%%%%%%%%%%%%%%%%%%%%%%%%%%%%
% Begin Figure
%%%%%%%%%%%%%%%%%%%%%%%%%%%%%%%%%%%%%%
\begin{figure}
\begin{center}
\includegraphics[angle=0,width=3.0in]{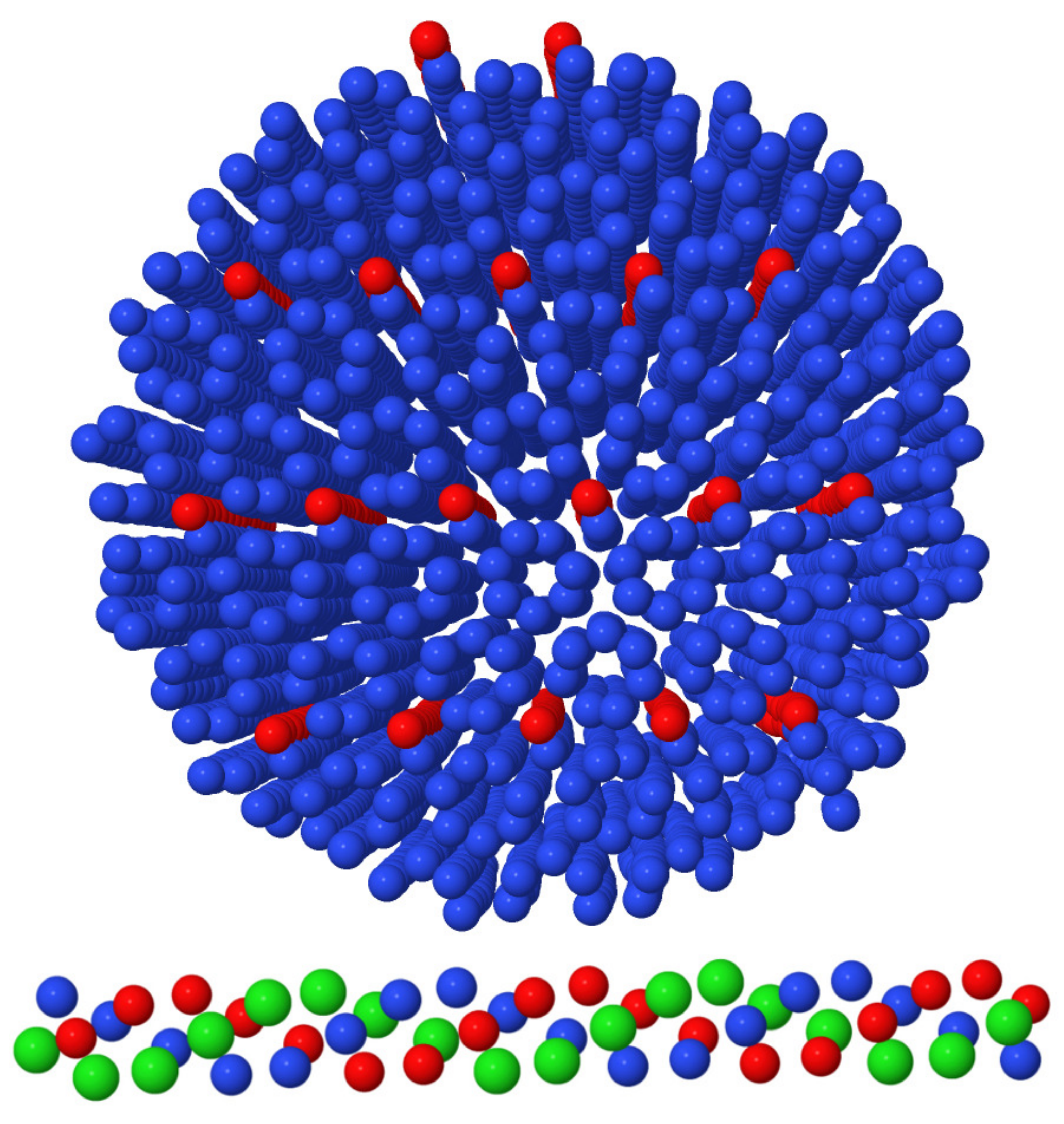}
\caption{
The upper figure shows a well-ordered extract 
from the structure formed in MD simulations after cooling
to nearly zero temperature at $\rho_o \sigma^3 = 2.904$. 
All particles are the same.
Particles located at the vertices of the guessed unit cells
are colored in red. 
Notice that the lines formed by particles organize into
columns and that each column is formed
by 7 lines of particles. 
The lower figure shows a particular column
from a direction perpendicular to the axis of the column.
It shows that it is also possible to assume that each column is formed by three helical coils.
There are 7 particles in the pitch of every coil. 
}\label{fig:7memeber-01}
\end{center}
\end{figure}
%%%%%%%%%%%%%%%%%%%%%%%%%%%%%%%%%%%%%%
% End Figure
%%%%%%%%%%%%%%%%%%%%%%%%%%%%%%%%%%%%% 

\section{Results for particular densities}\label{sec:results} 
 
Table \ref{table:MD-summary-1} summarizes the data obtained in our MD simulations.
More detailed descriptions of the results at the studied densities are given below.

\subsection{Densities $\rho_o \sigma^3 = 1.68$ and $\rho_o \sigma^3 = 1.84$}

The PDFs of the structures obtained through crystallization of liquids, 
after cooling to nearly zero temperature, are shown in Fig.\ref{fig:pdfs-01}(a,b). 
These figures also show the PDFs calculated on the Face-Centered Cubic (FCC) 
lattices that were produced by running LAMMPS on the FCC lattices without defects. 
Note that the temperatures for the different curves are different. 
In our view, it is clear from Fig.\ref{fig:pdfs-01}(a,b)
that liquids at these densities crystallize into highly defective FCC lattices.

Note, in the following Fig.\ref{fig:potlatt1x}, that at 
the densities that we discuss in this subsection the FCC structure 
has the smallest potential energy between the all considered structures.

\subsection{Densities $\rho_o \sigma^3 = 2.11$ and $\rho_o \sigma^3 = 2.40$ }

The PDFs of the structures obtained from crystallization of liquids, after cooling 
to nearly zero temperature, are shown in Fig.\ref{fig:pdfs-01}(c,d). 
The figures also show the PDFs calculated on the Body-Centered Cubic (BCC) 
lattices that were produced by running LAMMPS on the BCC lattices created without defects. 
Note that the temperatures for the different curves are different. 
In our view, it is clear from Fig.\ref{fig:pdfs-01}(c,d) that the 
liquids at these densities crystallize into the defective BCC lattices.

Note in Fig.\ref{fig:potlatt1x} that at the densities 
that we discuss in this subsection the BCC
structure has the smallest potential energy between all considered structures.

\subsection{Density $\rho_o \sigma^3 = 2.904$}

The liquid at this density exhibited rather significant resilience against crystallization, i.e.,
in order to observe crystallization it was necessary to perform rather long MD runs at 
temperatures at which liquid's dynamics is already slow.
The crystal structure observed at this density is, probably, the most complex between 
all of the observed structures.  
For this reason we describe our results in details.

We melted the BCC lattice consisting of 18522 particles in a cubic simulation box at $T=15\cdot 10^{-3}$. 
After equilibration the liquid was cooled using some cooling rate with longer relaxation times at lower 
temperatures when the diffusion is already slow. We observed crystallization, using 
the dependence of the PEpP on time, at $T \approx 2.5\cdot 10^{-3}$. 

If thus obtained structure is heated then it melts at 
$10^3T \approx 4.38\cdot 10^{-3}$. 
Then the structure obtained through crystallization was further cooled to nearly zero temperature. 

The PDF of the thus obtained structure at nearly zero temperature is shown in Fig.\ref{fig:pdfs-01}(e).
A snapshot of a well-ordered extract from the system is shown in Fig.\ref{fig:7memeber-01}. 
Visual analysis of this extract allowed to guess a triclinic unit cell with 16 particles
in it.

\begin{center}
\begin{table}
\begin{tabular}{| c | c | c | c | c |} \hline
Wyckoff site & $x$ & $y$ & $z$ & $10^2 U_i$  \\\hline
1st (8f) site & $0.1623$ & $0.1174$ & $-0.4496$ & $51.83$  \\\hline
2nd (8f) site & $-0.1720$ & $0.1312$ & $0.2514$ & $49.97$  \\\hline
3rd (8f) site & $0.0089$ & $0.4649$ & $0.1142$  & $53.76$  \\ \hline
4th (8f) site & $0.3723$ & $0.2964$ & $0.1157$  & $52.06$ \\\hline
\end{tabular}
\caption{
The coordinates of the four different non-special Wyckoff (8f) sites in the monoclinic unit cell
in terms of its edge vectors. 
The lengths of the edge vectors and the angles between them are:
$a = b = 2.1167$, $c=2.6205$, $\alpha = \gamma = 90^{\circ}$, $\beta = 110.20^{\circ}$.
The last column shows the half of the interaction energy of the site
in the ideal lattice at zero temperature.
}
\label{table:C2c-tbl}
\end{table}  
\end{center}

The guessed structure without defects was used as input for the molecular dynamics LAMMPS
program in order check the stability of the guessed structure and to refine the positions of particles within the unit cells.
The guessed structure exhibited stability at temperatures $T < 6.0\cdot 10^{-3}$.
Then the FINDSYM program was used to classify the unit cell(s) from the refined structure 
at zero temperature \cite{findsym1,findsym2}.
The highest symmetry solution found by FINDSYM 
at reasonably small values of the tolerance 
for the lattice parameters and particles coordinates  
is the monoclinic unit cell that belongs to the
$C2/c$ ($\#15$) crystallographic space group. 
According to the found solution, there are 32 particles in the unit cell 
occupying four different non-special Wyckoff (8f) sites. 
We found that particles occupying different Wyckoff sites have
different potential energies. 
The parameters of the classified unit cell are presented in Table \ref{table:C2c-tbl}.

We would like to note that often the FINDSYM programs finds a lower symmetry solution, i.e.,
the monoclinic lattice $Cc$ (space group $\#9$) with particles occupying eight different 
(4a) Wyckoff sites. 
However, we decided to describe here in details the solution with the highest symmetry.

For the discussed structure we did not attempt to perform the detailed optimization of the
lattice structural parameters, as their changes are likely to lead to the adjustments 
of the positions of the non-special Wyckoff sites. 
For this reason, the structural optimization of the observed structure appears to be 
a complicated task that deserves a separate investigation.

We observed the formation 
of the ``7-columns ring structures" in 5 independent MD runs. 
Four runs were made with the systems of particles 
containing $4 \cdot 15 \cdot15 \cdot 15 =13500$ particles and one
MD run was made with the system containing $2 \cdot 21 \cdot 21 \cdot 21 = 18522$ 
particles (this run has been started from melting the BCC lattice).
In all cases we observed crystallization at $T/\epsilon = 0.25$.
In order to observe crystallization it was necessary to wait 
for as long as $10^7$ MD steps ($1$ MD step corresponds to $0.001\tau$).

A characteristic feature of our guessed structure is that not all 
particles in it have identical environments, as the last column in Table \ref{table:C2c-tbl} shows.
The possibility of formation of the structures in which not all particles have identical local environments, 
while all particles are the same, has been discussed in Ref.\cite{Stillinger20131}. 
 
Finally, we would like to note that the columnar organization of the observed crystal structure
resembles (from a local perspective) the organization of particles in some
columnar quasicrystals \cite{Ryltsev20151,Ryltsev20171,Damascento20171,Dzugutov19931}.

In our view, further investigations of the observed columnar structure 
and of the mechanism of its formation are of interest.

\subsection{Densities $\rho_o\sigma^3 = 3.250,\;\;3.352,\;\;3.900,\;\;4.1232,\;\;4.1600$.\\
Observation of the $Ia\bar{3}d$ crystal structure.}\label{sec:LLL-lattice-1}

%%%%%%%%%%%%%%%%%%%%%%%%%%%%%%%%%%%%%%
% Begin Figure
%%%%%%%%%%%%%%%%%%%%%%%%%%%%%%%%%%%%%%
\begin{figure}
\includegraphics[angle=0.0,width=3.3in]{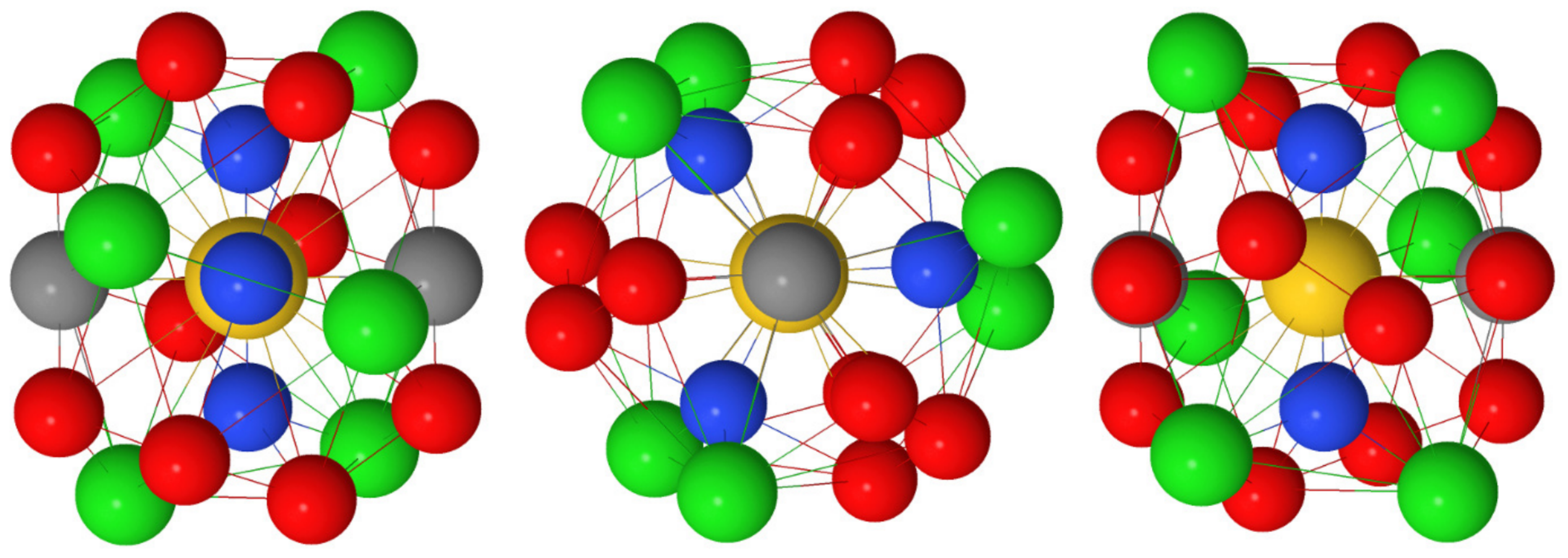}
\caption{The structure of the local environment of \emph{every particle}.
All particles are the same despite different colors used in the figure. 
Every central particle (\emph{yellow}) has 
3 first nearest neighbors (blue), 
2 second nearest neighbors (\emph{grey}), 
12 third nearest neighbors (\emph{red}), and 
6 forth nearest neighbors (\emph{green}).
The distances from the central particle to 
its neighbors are shown in table \ref{table:geom-vs-arith}.
}\label{fig:shell4neighbrs}
\end{figure}
%%%%%%%%%%%%%%%%%%%%%%%%%%%%%%%%%%%%%%
% End Figure
%%%%%%%%%%%%%%%%%%%%%%%%%%%%%%%%%%%%%

%%%%%%%%%%%%%%%%%%%%%%%%%%%%%%%%%%%%%%
% Begin Figure
%%%%%%%%%%%%%%%%%%%%%%%%%%%%%%%%%%%%%%
\begin{figure}
%\begin{center}
\includegraphics[angle=0,width=2.5in]{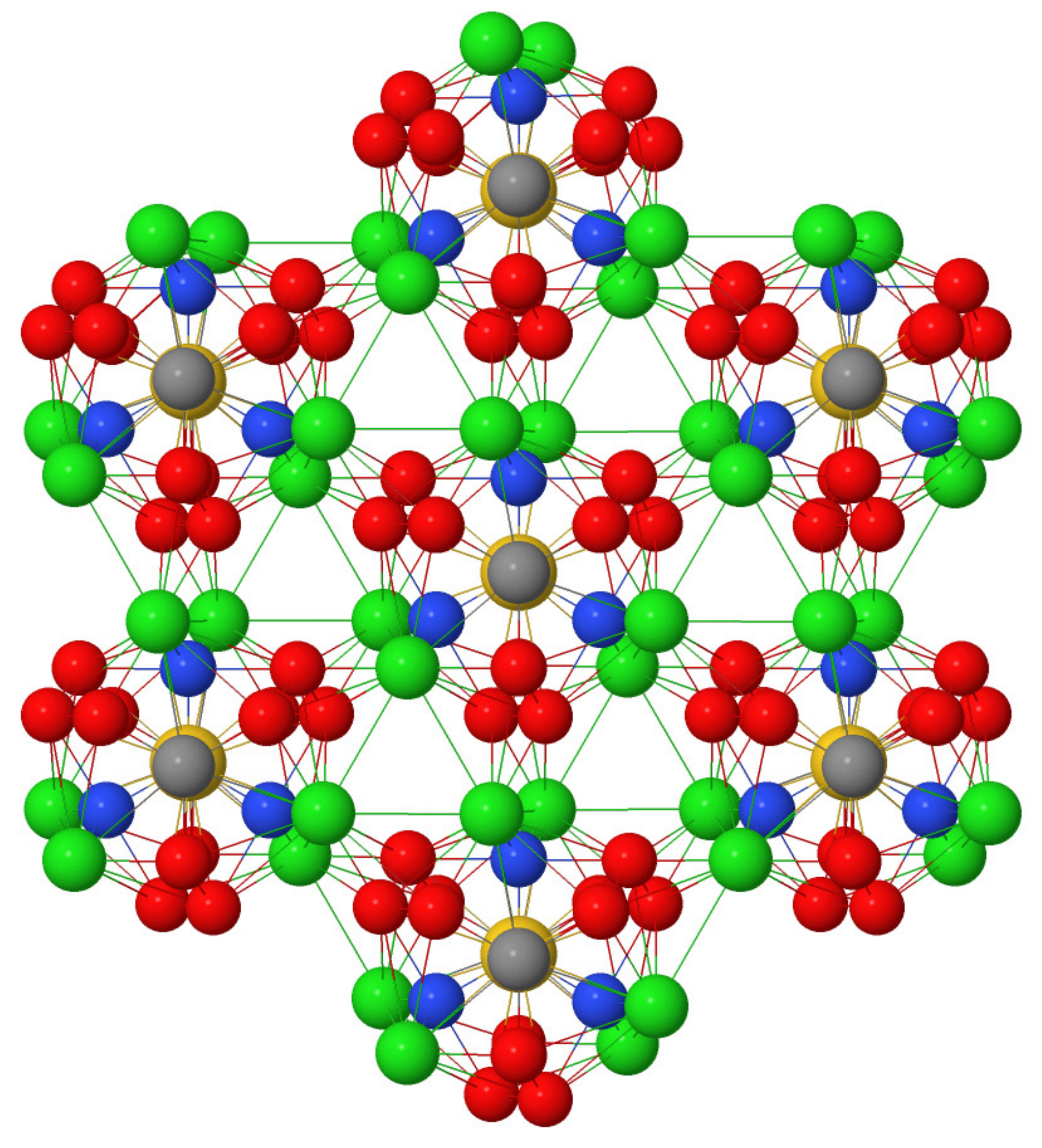}
\caption{
The organization of the ``atomic shell units" into a layer of the structure.
Note that around the central shell there are six empty ``triangles" into 
which additional particles should be placed at the proper heights. 
In the final structure all particles have identical environments.
}\label{fig:shell-units-organization}
%\end{center}
\end{figure}
%%%%%%%%%%%%%%%%%%%%%%%%%%%%%%%%%%%%%%
% End Figure
%%%%%%%%%%%%%%%%%%%%%%%%%%%%%%%%%%%%% 

Initially we studied the behavior of particles at density $\rho_o \sigma^3 =3.352$.
After crystallization, we performed relatively long relaxation runs in order to obtain a better quality PDF. 
These long runs indeed helped in improving the quality of the PDF and correspondingly of the crystal structure. 
Further we visually analyzed the local atomic environments of several selected particles. 
We found that nearly all selected particles have similar environments. 
The idealized version of this environment is shown in Fig.\ref{fig:shell4neighbrs}. 
See also table \ref{table:geom-vs-arith}.

Organization of the neighbor environment of particles in Fig.\ref{fig:shell4neighbrs} 
suggests the organization of these units into the layered structure, with 
the structure of layers shown in Fig.\ref{fig:shell-units-organization}.
The striking fact from the structure organization presented in Fig.\ref{fig:shell-units-organization} 
is that all particles in the structure have the identical neighbor environments shown in Fig.\ref{fig:shell4neighbrs}.

The guessed structure presented in 
Fig.\ref{fig:shell-units-organization} allows reconstructing a unit cell in a hexagonal lattice. 
On the next step, thus guessed crystal structure was classified with the Findsym \cite{findsym1,findsym2} software.
Running Findsym with the input of atomic coordinates in the hexagonal unit cell with moderate values
of the ``tolerance" parameter yielded cubic $Ia\bar{3}d$ (space group $\#230$) lattice
with 16 particles per unit cell occupying the (16b) Wyckoff special positions. 
The coordinates of the particles in the cubic unit cell are given in table \ref{table:ceLLL}.

In order to verify the structural guess we generated the $Ia\bar{3}d$-crystal structure 
without defects at this density and used it as the initial configuration for the LAMMPS program. 
Thus we found that this structure remains stable up to the temperature $T=0.010$. 
We also found that there is essentially a perfect agreement between the PDF obtained 
through crystallization of the liquid and the PDF calculated on the $Ia\bar{3}d$-crystal 
structure, as shown in Fig.\ref{fig:pdfs-01}(f).

\begin{center}
\begin{table}
\caption{The numbers of the $n$th order neighbors and 
the distances to them in the $Ia\bar{3}d$-lattice formed at density $\rho_o \sigma^3 = 3.352$. 
In the table ``NBR" stands for ``neighbor".
}\label{table:geom-vs-arith}
\begin{tabular}{| c | c | c | c |} \hline
NBR order &  $r_{n}/\sigma$  & \# of NBR & Color of NBR in Fig.\ref{fig:shell4neighbrs}\\ \hline
$1$ & $ \approx 0.59$ & $3$  & Blue\\ \hline
$2$ & $ \approx 0.73$ & $2$  & Grey\\ \hline
$3$ & $ \approx 0.94$ & $12$ & Red\\ \hline
$4$ & $ \approx 1.03$ & $6$  & Green \\ \hline
\end{tabular}
\end{table}  
\end{center}
\begin{center}
\begin{table}
\caption{The coordinates of the particles in the cubic unit 
cell of the $Ia\bar{3}d$ lattice with particles located at the (16b)
Wyckoff special positions. It is assumed that the length 
of the side of the unit cell
is equal to 8 (eight).}\label{table:ceLLL}
\begin{tabular}{| c | c | c | c |} \hline
$(1,1,1)$ & $(3,3,3)$ & $(5,5,5)$ & $(7,7,7)$  \\ \hline
$(1,3,5)$ & $(1,7,3)$ & $(1,5,7)$ & $(3,7,5)$  \\ \hline
$(3,5,1)$ & $(7,3,1)$ & $(5,7,1)$ & $(7,5,3)$  \\ \hline
$(5,1,3)$ & $(3,1,7)$ & $(7,1,5)$ & $(5,3,7)$  \\ \hline
\end{tabular}
\end{table}  
\end{center}

Then we performed MD simulations at several other densities around $\rho_o \sigma^3 = 3.352$. 
These densities are listed in the title of this subsection. 
We found that at these densities the PDFs obtained through 
crystallizations of liquids appear to be qualitatively similar to the PDF 
obtained at density $\rho_o \sigma^3 = 3.352$, as can be seen from the comparison 
of the curves in  Fig.\ref{fig:pdfs-01}(f) and Fig.\ref{fig:pdfs-02}(a,b). 
Then we generated the $Ia\bar{3}d$-structures at these densities and performed 
MD simulations on them. 
These simulations showed the stability of the $Ia\bar{3}d$-structure at these densities.

Note in Fig.\ref{fig:potlatt1x} that at the densities 
that we discuss in this subsection the guessed $Ia\bar{3}d$ structure has 
the lowest potential energy between the all considered structures.

According to the chemical structure database, the $Ia\bar{3}d$ crystal structure has not been observed 
before in the single component atomic systems. 
At the same time, there are binary and ternary atomic compounds that form the $Ia\bar{3}d$ 
crystal structure. The $Ia\bar{3}d$ crystal structure is also known to form 
as the superstructure formed by the amphiphilic (for example, soap) 
molecules \cite{Luzzati1968,Seddon2006,IsraelachviliJ2011,Kutsumizu2012}. 
It is known that the amphiphilic molecules can organize into spherical 
micelles \cite{Luzzati1968,Seddon2006,IsraelachviliJ2011,Kutsumizu2012}.
Moreover, it has been suggested that the interaction between spherical 
colloids and emulsions can be modeled with the harmonic-repulsive pair potential \cite{Mohanty20141,Jorjadze20131}. 
In this context, our results suggest that the spherical micelles 
can organize into the $Ia\bar{3}d$ crystal structure, which on further increase 
in the concentration of the amphiphilic molecules ``polymerizes" into the two 
reticular systems of the tubes with the nodes forming the $Ia\bar{3}d$ 
structure \cite{Luzzati1968,Seddon2006,IsraelachviliJ2011,Kutsumizu2012}.   
The $Ia\bar{3}d$ structures also can be formed by 
liquid crystals and gels \cite{Tschierske2013,Cho20151}.
The $Ia\bar{3}d$ crystal structure also can 
be associated with gyroid minimal surfaces \cite{Cho20151}.
The $Ia\bar{3}d$ crystal structure has been described as the 
system composed of two interpenetrating nets in Ref.\cite{Wells19541,Lobanov-Lokshin}.

\subsection{Densities $\rho_0 \sigma^3 = 4.40$ and $\rho_0 \sigma^3 = 4.50$}

\begin{figure}
\begin{center}
\includegraphics[angle=0,width=3.3in]{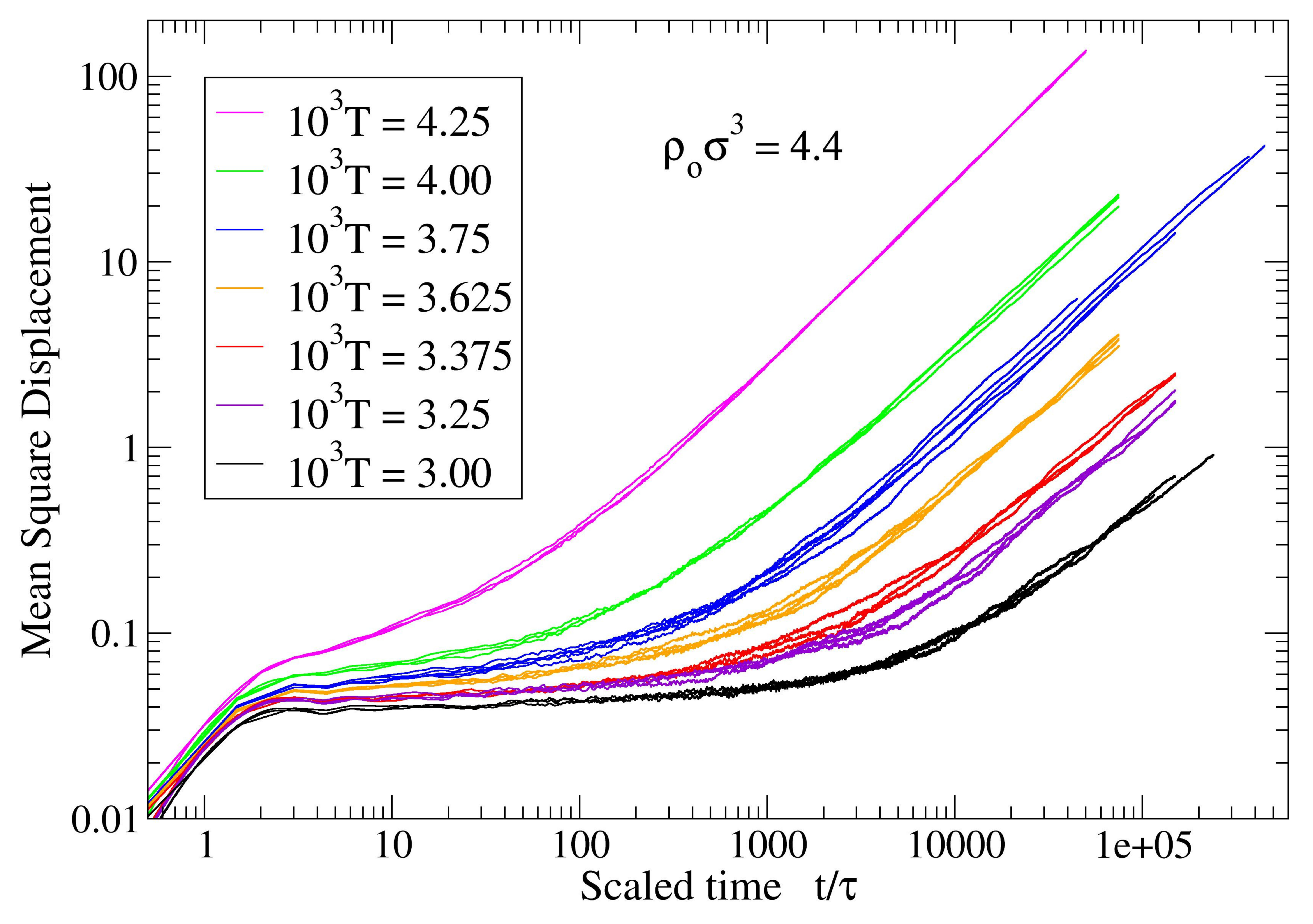}
\caption{The dependencies of the mean square particle's displacement on time
in MD runs at the selected temperatures. 
All data were obtained at the density $\rho_o \sigma^3 = 4.4$.
The different curves of the same color correspond to the several 
consecutive runs at the same temperature. 
The system at $10^3T = 3.75$ exhibits slow relaxation
(especially in the earlier runs). 
This relaxation is partially responsible for the observable differences in the
blue curves.
}\label{fig:msd-vs-time-1}
\end{center}
\end{figure}

\begin{figure*}
\begin{center}
\includegraphics[angle=0,width=7.0in]{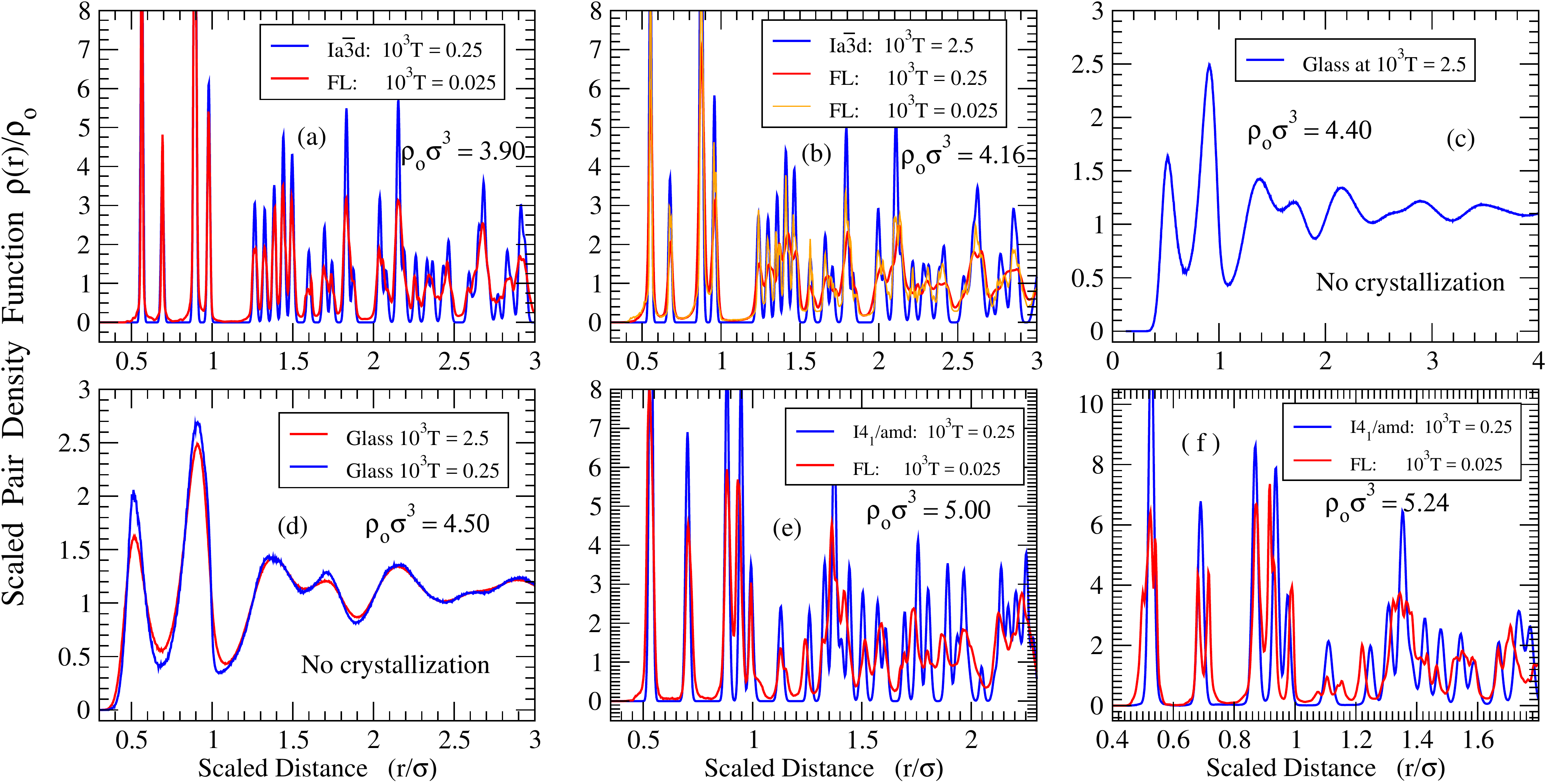}
\caption{The PDFs, $\rho(r)$, of the crystal structures obtained 
through crystallization from the liquid states and the PDFs of 
the corresponding guessed structures at the particles' densities 
shown in the panels. The ``FL" notation stands for ``From Liquid".
At the densities $\rho_0 \sigma^3 = 4.40$ and $\rho_0 \sigma^3 = 4.50$ we did not observe 
crystallization despite rather long MD runs 
(see Fig.\ref{fig:msd-vs-time-1}).
}\label{fig:pdfs-02}
\end{center}
\end{figure*}

At these densities we did not observe crystallization despite rather long molecular dynamics 
runs at a number of temperatures where the dynamics slows down and becomes very slow.
See Fig.\ref{fig:msd-vs-time-1}. The PDFs at these densities at very low temperatures are shown in Fig.\ref{fig:pdfs-02}(c,d).
Our visual analysis of the zero-temperature structures at these densities did not reveal the presence 
of a distinct structural pattern.

The stability of single component systems against crystallization is 
unusual because single component supercooled liquids usually readily 
crystallize \cite{HansenJP20061,Frenkel20021,Frenkel19961}.
In particular, several binary models have been intentionally 
developed to avoid crystallization and allow numerical investigations 
of the deeply supercooled model liquids \cite{Hansen19871,Kob19951}

However, recently there have been observations of the unusual stabilities of the single components 
liquids against crystallization \cite{Ryltsev20131,Damascento20171}. 
Our observations of the stability of the single component systems appear to 
be similar to the observations made in \cite{Ryltsev20131,Damascento20171}. 
At the same time, it appears to be of interest to note that the pair potential 
used in our study is noticeably simpler than the potentials studied in \cite{Ryltsev20131,Damascento20171}. 

%%%%%%%%%%%%%%%%%%%%%%%%%%%%%%%%%%%%%%
% Begin Figure
%%%%%%%%%%%%%%%%%%%%%%%%%%%%%%%%%%%%%%
%%%%%%%%%%%%%%%%%%%%%%%%%%%%%%%%%%%%%%
% End Figure
%%%%%%%%%%%%%%%%%%%%%%%%%%%%%%%%%%%%% 
%%%%%%%%%%%%%%%%%%%%%%%%%%%%%%%%%%%%%%
% Begin Figure
%%%%%%%%%%%%%%%%%%%%%%%%%%%%%%%%%%%%%%
%%%%%%%%%%%%%%%%%%%%%%%%%%%%%%%%%%%%%%
% End Figure
%%%%%%%%%%%%%%%%%%%%%%%%%%%%%%%%%%%%% 

\subsection{Densities $\rho_0 \sigma^3 = 5.0$ and $\rho_0 \sigma^3 = 5.24$. }

The PDFs of the crystal structures obtained from the liquid states 
at these densities are shown in Fig.\ref{fig:pdfs-02}(e,f).

The visual analysis of the crystal structure at $\rho_0 \sigma^3 = 5.0$ formed from 
the liquid with consequent cooling to zero temperature led us to the following 
guess of the minimal energy crystal structure.
The structure described below was optimized to correspond to the minimum value of the potential energy, 
i.e., the parameters $a$, $c$, $\gamma$, and $d$ were optimized to provide the minimum value of the potential energy.

Two opposite faces of the monoclinic unit cell are \emph{squares} with edges $a \approx 0.704$. 
Further, we assume that the translational unit vectors $\hat{T}_1$ and $\hat{T}_2$ are directed 
along the orthogonal edges of these squares. The projection of the third translational vector, $\hat{T}_3$, 
on the plane of $(\hat{T}_1,\hat{T}_2)$
forms  $45^{\circ}$ angles with $\hat{T}_1$ and $\hat{T}_2$. 
The angle that $\hat{T}_3$ forms with the plane of $(\hat{T}_1,\hat{T}_2)$, i.e., 
with the vector $(1/\sqrt{2})(\hat{T}_1+\hat{T}_2)$, is $\gamma \approx 58.361^{\hat{\circ}}$.
The length of the third translational vector is $c \approx 0.949$.
In the described structure the lengths of the projections, $p$, of the third translational 
vector on the first and second translational vectors are equal to half lengths of the first 
and second translational vectors ($p \approx a/2$).

Thus, in terms of the orthogonal Cartesian unit 
vectors $\hat{x}$, $\hat{y}$, $\hat{z}$, we write:
\begin{eqnarray}
&&\hat{T}_1 = \hat{x},\;\;\hat{T}_2 = \hat{y},\;\;\\
&&\hat{T}_3 =\tfrac{\sqrt{2}}{2}\cos(\gamma)\hat{x}+\tfrac{\sqrt{2}}{2}\cos(\gamma)\hat{y}+\sin(\gamma)\hat{z}.\;\;\;\;\;\;\;
\end{eqnarray}

Then, according to the visual analysis of the structure and correspondingly to the guess, inside 
the unit cell there is one additional atom whose position is described by the vector
\begin{eqnarray}
\vec{r}_2 = (a-d)\hat{T}_1 + (d)\hat{T}_2 + (c/2)\hat{T}_3,
\label{eq:second-atom-1}
\end{eqnarray} 
where $d \approx a/4$.

We found that the described guessed and optimized structure remains stable in MD simulations 
up to the temperature $10^3T = 9.38$, where it melts. No signatures of instability 
with respect to a transition into a different crystal structure have been observed on heating. 

We classified the guessed structure with the Findsym software \cite{findsym1,findsym2}. 
According to the Findsym solutions obtained for several small values of the tolerance 
parameter the guessed and optimized structure belongs to the tetragonal spatial 
group $I4_1/amd$ ($\#141$) with particles occupying (4b) Wyckoff positions. 
This structure is also known as the $A5$ structure of the $\beta Sn$.  
The possibility of formation of this structure has been assumed for 
the Hertzian potential in \cite{Prestipino20091} and it has been found that at some
pressures this structure is indeed more stable than the other considered structures. 
This structure also has been observed in simulations with pair potentials 
more complex than the potential used in the present study \cite{Prestipino20101,Malescio20111}.

The parameters of the optimized $I4_1/amd$ unit cell 
that lead to the lowest value of the potential energy are 
$a=b=0.7038$, $c=1.6154$,
$\alpha=\beta=\gamma = 90^{\circ}$. 
The coordinates of the particles occupying 
the (4b) Wyckoff positions inside the unit cell, 
in terms of the unit cell edge vectors, are
$(0,1/4,3/8),(1/2,3/4,7/8),(0,3/4,5/8),(1/2,1/4,1/8)$. 

We note that the dependence of the potential energy
function on the values of the parameters $a$ and $b$ near 
their optimized values is rather weak, i.e., the bottom of the
potential energy surface is rather flat.

According to Ref.\cite{LuZY20111} the equilibrium crystal structure 
at density $\rho_o \sigma^3 = 5.0$ is the diamond structure. 
In principle, it is possible to think about the structure that we observed 
as about a strongly distorted diamond structure. 
In our guessed structure every particle also has 4 nearest neighbors 
and all these neighbors are at the distance $0.535$ from the ``central" chosen particle. 
The difference with the diamond structure is in the values of the angles. 
In the diamond structure all 6 angles associated with the ``central" atom are equal
to $109.5^{\circ}$. In our guessed structure 2 angles 
are equal to $82.1^{\circ}$ (one can think about them as about two opposite angles), 
while the other 4 angles are equal to $124.6^{\circ}$. 
Thus the structure that crystallized from the liquid in 
our simulations is, in some sense, a distorted diamond structure.

For the density $\rho_0 \sigma^3 = 5.24$ the guessed structure, 
according to the visual analysis, is similar to the one at $\rho_0 \sigma^3 = 5.0$. 
The values of the parameters for the density $\rho_0 \sigma^3 = 5.24$ are 
the following: $a =0.691$, $c = 0.934$, $\gamma = 58.5645^{\circ}$, $d = a/4$. 
In terms of making comparison to the diamond structure 
for $\rho_0 \sigma^3 = 5.24$ we have: the distance to the 4 nearest neighbors is $0.530$, 
while the values of the angles are $81.7^{\circ}$ (2 angles) and 
$124.9^{\circ}$ (4 angles).

The symmetry of the solution that the Findsym software \cite{findsym1,findsym2} 
found for the density $\rho_o \sigma^3 = 5.24$ is the same as for the density $\rho_o \sigma^3 = 5.00$.
The optimized parameters of the unit cell are: $a = b = 0.6910$, while $c=1.5987$.

The comparisons of the PDFs of the crystal structures obtained through the crystallization 
of liquids with the PDFs calculated on the ideal crystal structures are shown in Fig.\ref{fig:pdfs-02}(e,f). 
It follows from Fig.\ref{fig:pdfs-02}(e) that our guess is quite good for 
the density $\rho_0 \sigma^3 = 5.0$ -- at least for the first six peaks. 
On the other hand, in Fig.\ref{fig:pdfs-02}(f) there is a splitting of 
the second peak in the PDF of the crystal structure obtained from the liquids state. 
Our guessed model, despite being stable on heating, does not capture this splitting. 
We were not able to come up with a better structural guess and thus, for now, 
we leave the matter in its current state.

Note in Fig.\ref{fig:potlatt1x} that at the densities 
$\rho_0 \sigma^3 = 5.0$ and $\rho_0 \sigma^3 = 5.24$ the guessed $I4_1/amd$ or $A5$ 
structure has the lowest value of the potential energy between all of the considered structures.

According to Ref.\cite{LuZY20111} at the considered densities the diamond structure should be stable. 
We performed MD simulations on the diamond structure at 
density $\rho_0 \sigma^3 = 5.0$ for several temperatures. 
According to our results the diamond structure at this density is indeed stable.
However, we also found that the PEpP of our guessed 
structure ($u_{distorted\;\;diamond}=0.6709$) is lower than 
the PEpP of the diamond structure ($u_{diamond} = 0.6722$).

\subsection{Density $\rho_0 \sigma^3 = 6.088$}

First of all, we note that in order to observe crystallization at this density it was 
necessary to perform rather long simulation runs at the observed ``crystallization" temperature.  
  
Visual analysis of the structure obtained by crystallization from the liquid state clearly 
suggests the presence of a crystal motif which leads us to the following guess of the crystal structure. 

%%%%%%%%%%%%%%%%%%%%%%%%%%%%%%%%%%%%%%
% Begin Figure
%%%%%%%%%%%%%%%%%%%%%%%%%%%%%%%%%%%%%%
\begin{figure}
\begin{center}
\includegraphics[angle=0,width=3.2in]{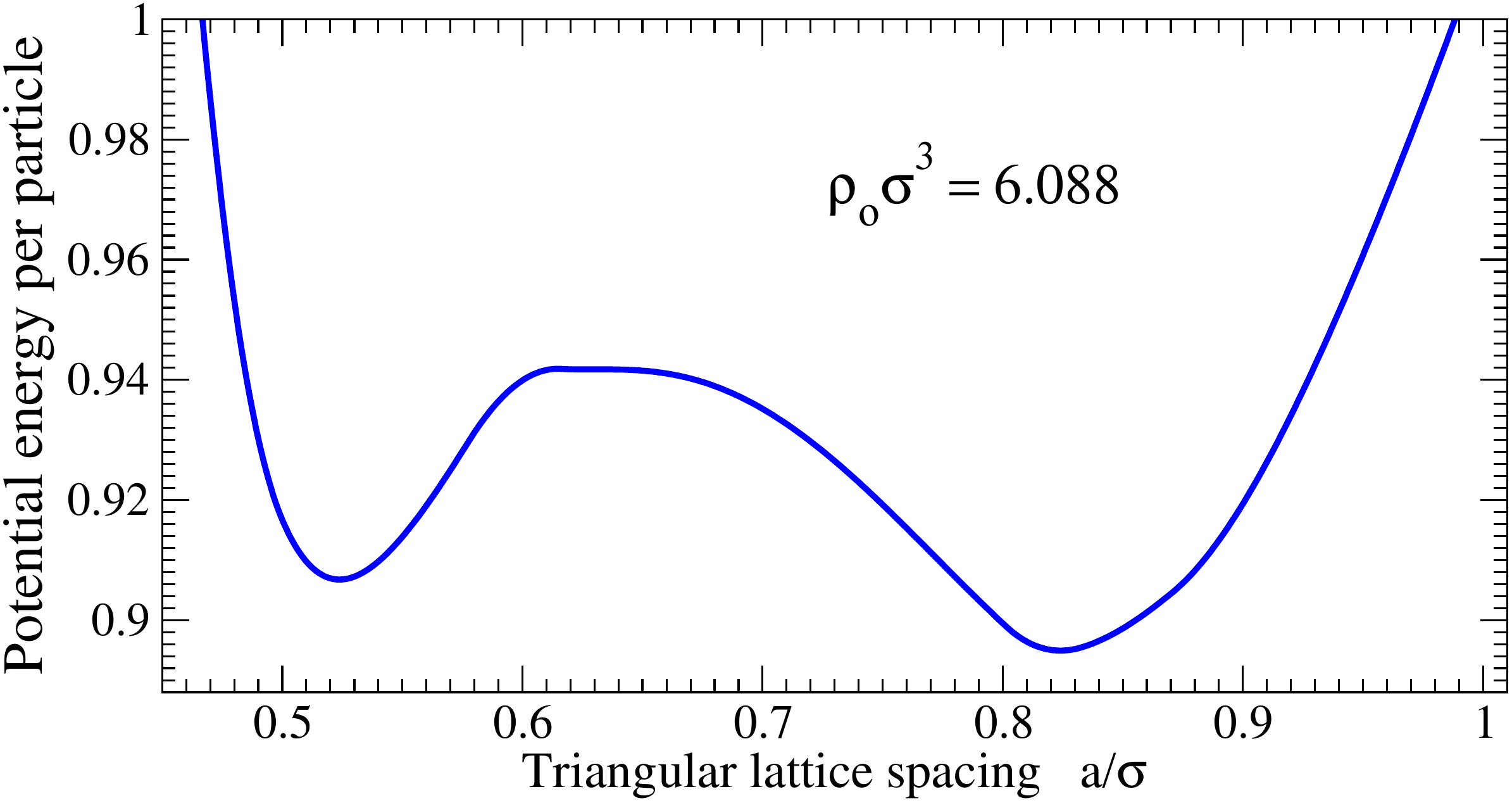}
\caption{
The potential energy per particle (PEpP) of the crystal structure formed by two interchanging 
triangular lattices at density $\rho_o \sigma^3 = 6.088$ as a function of the triangular lattice spacing, $a$. 
The required value of the density, $\rho_o$, can be achieved by a suitable choice of the triangular 
lattice spacing, $a$, and the spacing between the triangular lattices, $c/2$, i.e.: $c = (4/\sqrt{3})/(\rho_o a^2)$. 
Note the presence of two minima  in the potential energy curve. 
These two minima correspond to quite different values of $a$. 
The energy difference between the two minima is $\approx 0.012$. 
This energy difference, together with the height of the barrier between the minima,
$\approx 0.035$, should influence the crystallization process. 
It is possible to choose such value of the density at which the energies of the two minima 
are almost the same, $(\rho_o\sigma^3 \approx 5.624)$, for the considered lattice. 
This, of course, does not preclude the situation when some other lattice 
provides even lower value for the potential energy. 
}\label{fig:pdfs-03}
\end{center}
\end{figure}
%%%%%%%%%%%%%%%%%%%%%%%%%%%%%%%%%%%%%%
% End Figure
%%%%%%%%%%%%%%%%%%%%%%%%%%%%%%%%%%%%% 

%%%%%%%%%%%%%%%%%%%%%%%%%%%%%%%%%%%%%%
% Begin Figure
%%%%%%%%%%%%%%%%%%%%%%%%%%%%%%%%%%%%%%
\begin{figure*}
\begin{center}
\includegraphics[angle=0,width=7.0in]{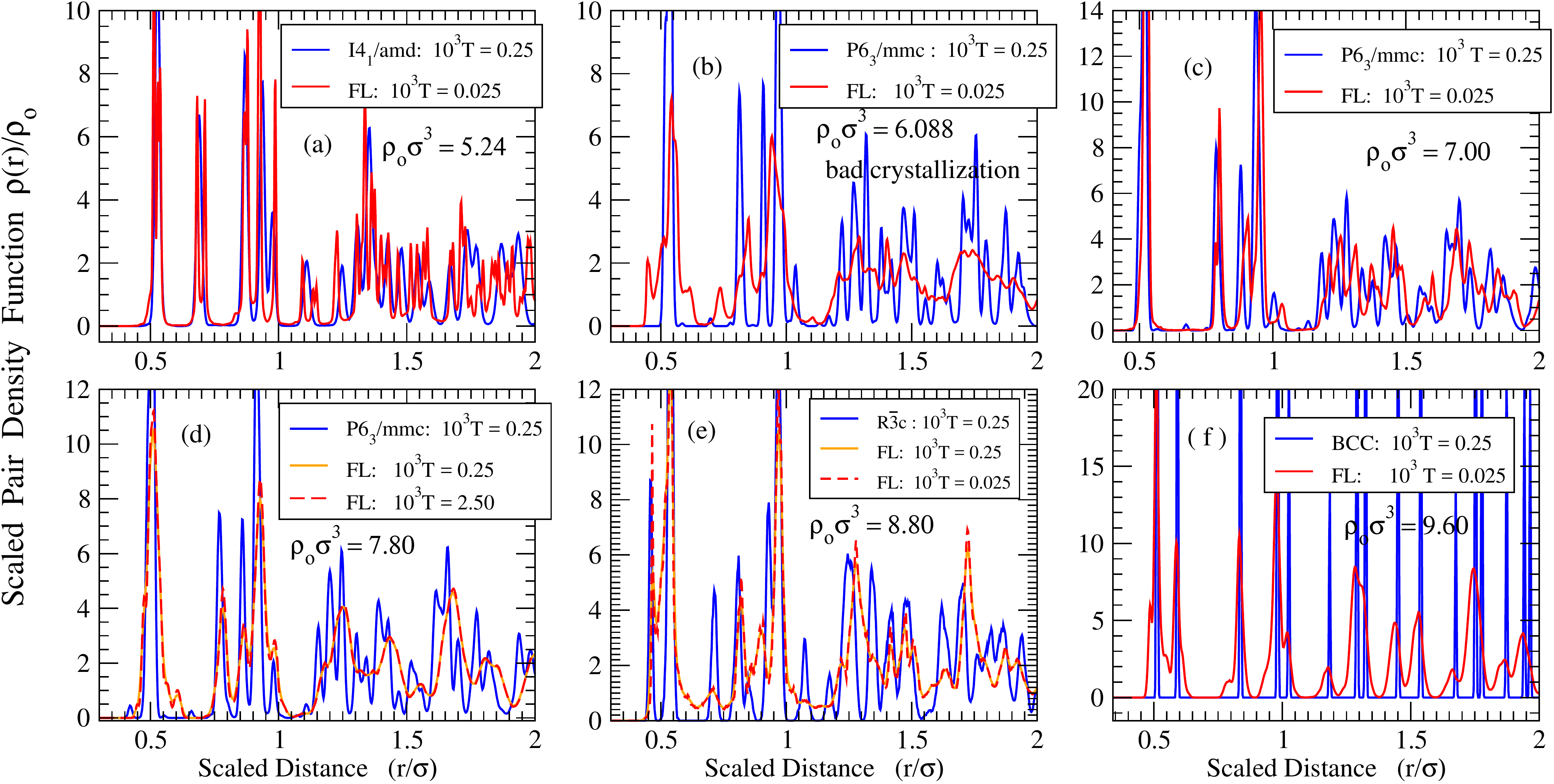}
\caption{
The PDFs, $\rho(r)$, of the crystal structures obtained 
through crystallization from the liquid states and the PDFs of 
the corresponding guessed structures at the particles' densities 
shown in the panels. The ``FL" notation stands for ``From Liquid".
}\label{fig:pdfs-03}
\end{center}
\end{figure*}
%%%%%%%%%%%%%%%%%%%%%%%%%%%%%%%%%%%%%%
% End Figure
%%%%%%%%%%%%%%%%%%%%%%%%%%%%%%%%%%%%% 

The guessed structure consist of two triangular lattices alternating along the $\hat{z}$-axis with 
the mutual orientation similar to the one observed in the Hexagonal Closed Packed (HCP) crystal lattice. 
However, the guessed structure is not the HCP lattice because in the HCP lattice there is a relation, $(c/a) = \sqrt{8/3} \approx 1.633$, 
between the lattice spacing of the triangular lattices, $a$, and the spacing 
between the nearest triangular lattices, $c$. 
In our guessed lattice this relation does not hold.
The optimization of the guessed structure to achieve the minimum value 
of the potential energy leads to the values $a = 0.824$ and $c = 0.559$, i.e., 
we have $(c/a) \approx 0.678$. Thus, the guessed structure, 
despite the fact that it consists of two alternating triangular lattices, 
is not the HCP lattice.

As we mentioned above, the motif of the two alternating triangular lattices can 
be clearly observed in the highly defective crystal structure that formed from the liquid state. 
However, the PEpP of the optimized (defect-free) guessed structure ($u_{optimized}\approx 0.8949$) 
is (slightly) higher than the PEpP of the highly defective structure formed 
from the liquid state ($u_{from\;liquid} \approx 0.8946$). 
At present, we can think of two possible explanations for this situation.\\
1) One possibility is that our guess of the structure
is simply incorrect. In any case, we were not able to make a better guess. 
Moreover, we found that the guessed structure, when
we used it as a starting configuration for the MD program,
is stable in a wide range of temperatures, 
as can be seen in Table \ref{table:MD-summary-1}.\\ 
2) Another possibility is that the system at this overall density might achieve 
the lower value of the potential energy through phase separation into regions 
with different densities (with the same or different crystal structures).
Our analysis of the ``crystal" structure formed from the liquid state does 
not allow us to make more definite statements with respect to this
possibility.  

The comparison of the PDF calculated from the structure 
obtained through the crystallization of liquid with the PDF calculated
from the guessed structure is shown in Fig.\ref{fig:pdfs-03}(b).

We classified the guessed and optimized structure with the 
Findsym software \cite{findsym1,findsym2}. The Findsym provided the 
same solutions for several different values of the tolerance parameter. 
The solution is the hexagonal spatial group $P6_{3}/mmc$ 
(spatial group $\#194$) with particles occupying (2c) Wickoff positions.
The parameters of the classified unit cell are: 
$a=b = 0.824$, $c=1.118$,
$\alpha=\beta = 90^{\circ}$, $\gamma = 120^{\circ}$. 
The fractional coordinates of the particles inside the unit cell, 
in terms of the unit cell edge vectors, are
$(2/3,1/3,3/4)$ and $(1/3,2/3,1/4)$.
 
Note in Fig.\ref{fig:potlatt1x} that at density 
$\rho_0 \sigma^3 = 6.088$ the guessed $P6_{3}/mmc$
structure has the lowest value of the potential energy between 
all of the considered structures.

\subsection{Densities $\rho_0 \sigma^3 = 7.00$ and $\rho_0 \sigma^3 = 7.80$ }

Consider the PDFs calculated on the crystal structures obtained through the crystallization 
of liquids in panels (b,c,d) of Fig.\ref{fig:pdfs-03} and 
note that qualitatively they look similar.
This observation suggests that the ideal crystal structures at these densities are formed 
by two alternating triangular lattices.
Visual analysis of the crystal structures obtained through the crystallization of liquids 
at these densities supports this assumption.

The optimized values of the lattice parameters for density  
$\rho_0 \sigma^3 = 7.00$ are $a \approx 0.800$ and $c \approx 0.5155$.

The optimized values of the lattice parameters for density  
$\rho_0 \sigma^3 = 7.80$ are $a \approx 0.779$ and $c \approx 0.488$.

Note in Table \ref{table:MD-summary-1} that for the densities discussed in this subsection 
the values of the PEpP of the guessed structures without defects are lower than the values of the
PEpP of the structures obtained from the liquid state.

The MD simulations of the guessed structures demonstrated their stability. 
The comparisons of the PDFs calculated on the guessed structures with the PDFs calculated 
on the crystal structures obtained from the liquid states suggest that our guessed 
structures might indeed correspond to the ground state structures at these densities.
Note in Fig.\ref{fig:potlatt1x} that at the densities 
$\rho_0 \sigma^3 = 7.00$ and $\rho_0 \sigma^3 = 7.80$ the guessed $P6_{3}/mmc$ structure 
has the smallest potential energy between the all considered structures. 

The classifications of the structures at these densities with the Findsym software \cite{findsym1,findsym2} lead 
to the same solutions that were obtained for the density $\rho_o\sigma^3 = 6.088$ with 
the adjusted values of the lattice edge lengths.

\subsection{Density $\rho_0 \sigma^3 = 8.80$}

The PDF calculated on the structure obtained by crystallization of the liquid is shown in Fig.\ref{fig:pdfs-03}(e). 
Note that thus obtained PDF does not have well-defined peaks in comparison, for example, with Fig.\ref{fig:pdfs-03}(d).

\begin{figure}
\begin{center}
\includegraphics[angle=0,width=2.3in]{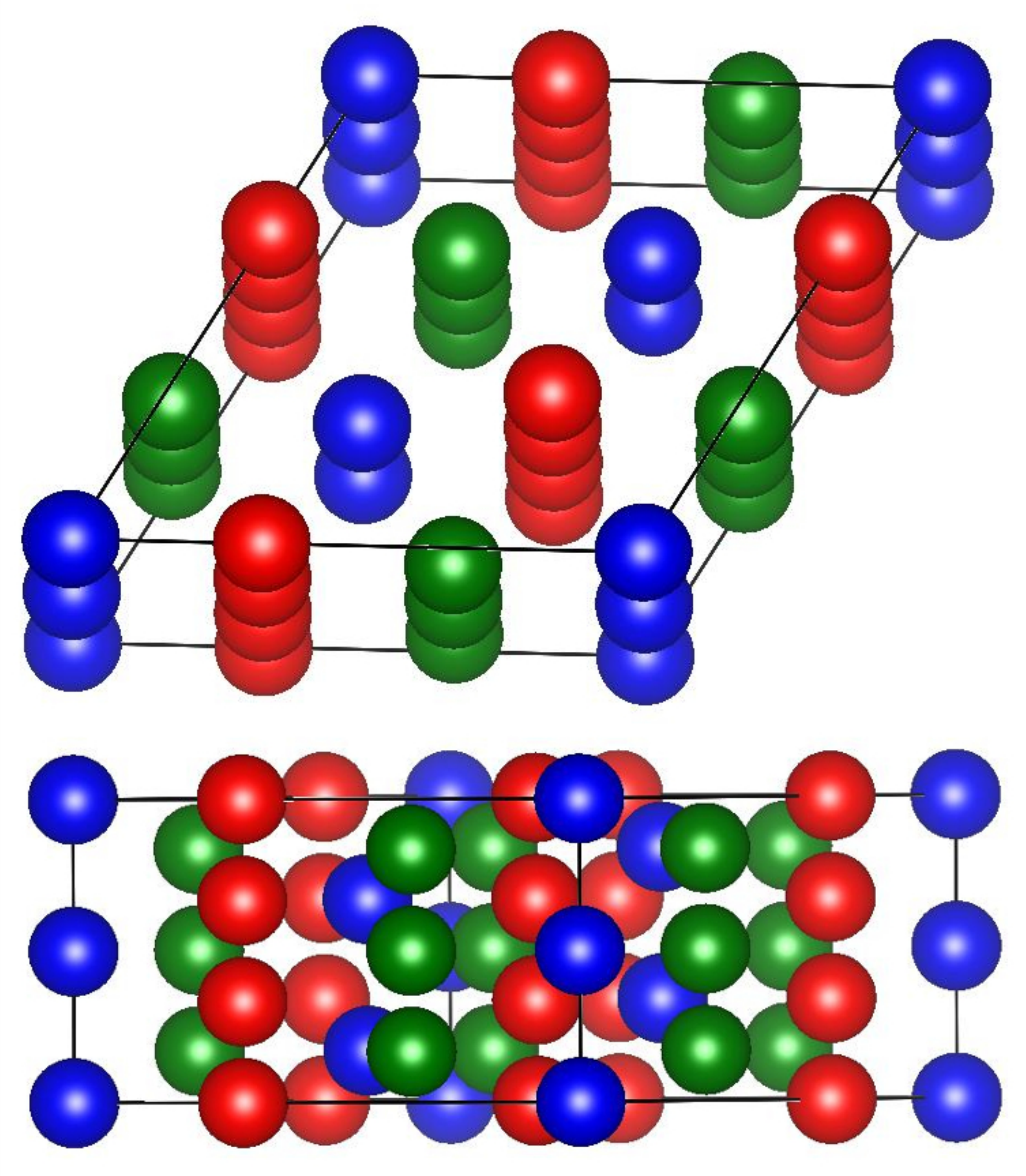}
\caption{
Two views on the guessed unit cell at the density $\rho_o\sigma^3 = 8.8$.
All particles are the same. Different colors were used to illustrate the structure.
In the shown guessed structure the environments blue particles are not the same 
as the environments of the red and green particles.
Therefore, the potential energies of all red and green particles are the same (1.437), 
but they are different from the energies of the blue particles (1.450).
}\label{fig:3view-8x8}
\end{center}
\end{figure}

Our visual analysis of the structure led us to the guessed crystal structure with the following translational vectors:
\begin{eqnarray}
\vec{T}_1 = a\hat{x},\;\;\vec{T}_2 = \left(\tfrac{1}{2}\right)a\hat{x} + \left(\tfrac{\sqrt{3}}{2}\right)a\hat{y},\;\;
\vec{T}_3 = c\hat{z},
\label{eq:T1T2T3rho8x8-1}
\end{eqnarray} 
where
\begin{eqnarray}
a = \sigma\cdot 4\left(\tfrac{3}{\rho_o\sigma^3}\right)^{1/3},\;\;c = \sigma\cdot 4\left(\tfrac{1/\sqrt{3}}{\rho_o\sigma^3}\right)^{1/3}.
\label{eq:T1T2T3rho8x8-2}
\end{eqnarray}
According to our guess there are 24 basis particles inside the unit cell defined by the translational vectors (\ref{eq:T1T2T3rho8x8-1}).
The coordinate of a particle $i$ inside the unit cell can be 
represented as $\vec{r}_i = b_1(i)\vec{T}_1 + b_2(i)\vec{T}_2 + b_3(i)\vec{T}_2$, where $b_n(i)$ are the fractional coordinates 
of the particle $i$ in terms of the translational vectors.
The fractional coordinates of all 24 particles in the guessed and optimized structure are given in Table \ref{table:fraccoord1}.
The guessed unit cell is also shown in Fig.\ref{fig:3view-8x8}. 
One can think of the guessed structure as being composed of five triangular lattices.
\begin{center}
\begin{table}
\small
\caption{The fractional coordinates of the basis particles (in terms of the translational vectors)
inside the unit cell of the guessed structure at the density $\rho_o\sigma^3 = 8.8$.
The first four lines give the coordinates of the 12 particles. The second four lines give
the coordinates of another 12 particles. The coordinates of the particles $1-6$ in the table
give the coordinates of the {\it blue} particles in the unit cell shown in Fig.\ref{fig:3view-8x8}.
The coordinates of the particles $7-15$ and $16-24$ in the table describe correspondingly the positions 
of the {\it green} and  {\it red} particles in Fig.\ref{fig:3view-8x8}.
}\label{table:fraccoord1}
\begin{tabular}{| c | c | c | c | c | c | c | c | c | c | c | c | c |}      \hline
$i\rightarrow$  &$1$& $2$       &$3$  & $4$ &$5$  & $6$ & $7$ & $8$ &$9$  & $10$&$11$ &$12$     \\ \hline
$b_1$           &$0$&$0$ &$1/3$&$1/3$&$2/3$&$2/3$   &$1/3$&$1/3$    &$1/3$&$0$  &$0$  &$0$                \\ \hline
$b_2$           &$0$&$0$ &$1/3$&$1/3$&$2/3$&$2/3$   &$0$  &$0$      &$0$  &$2/3$&$2/3$&$2/3$              \\ \hline
$b_3$           &$0$&$1/2$ &$1/6$&$2/3$&$1/3$&$5/6$ &$0$  &$1/3$    &$2/3$&$0$  &$1/3$&$2/3$            \\ \hline
$i\rightarrow$ &$13$ &$14$ &$15$ &$16$ &$17$ &$18$ &$19$ &$20$   &$21$&$22$&$23$&$24$     \\ \hline
$b_1$          &$2/3$&$2/3$&$2/3$&$2/3$&$2/3$&$2/3$&$0$  &$0$  &$0$  &$1/3$&$1/3$&$1/3$      \\ \hline
$b_2$          &$1/3$&$1/3$&$1/3$&$0$&$0$  &$0$  &$1/3$&$1/3$&$1/3$&$2/3$&$2/3$&$2/3$      \\ \hline
$b_3$          &$0$  &$1/3$&$2/3$&$1/6$&$1/2$&$5/6$&$1/6$&$1/2$&$5/6$&$1/6$&$1/2$&$5/6$    \\ \hline
\end{tabular}
\end{table}  
\end{center}

%%%%%%%%%%%%%%%%%%%%%%%%%%%%%%%%%%%%%%
% Begin Figure
%%%%%%%%%%%%%%%%%%%%%%%%%%%%%%%%%%%%%%
%%%%%%%%%%%%%%%%%%%%%%%%%%%%%%%%%%%%%%
% End Figure
%%%%%%%%%%%%%%%%%%%%%%%%%%%%%%%%%%%%% 

The comparison of the PDF calculated on the guessed structure with the PDF calculated on 
the structure obtained by crystallization from the liquid state is shown in Fig.\ref{fig:pdfs-03}(e). 
We found that the guessed structure remains stable on heating until melting at $T\approx 8.25\cdot10^{-3}$.

The classification of the guessed and optimized structure with the 
Findsym software \cite{findsym1,findsym2} lead to the Hexagonal space group $R\bar{3}c$ ($\#167$) 
with 24 particles in the unit cell occupying Wyckoff special positions (6a) and (18e).

Note in Fig.\ref{fig:potlatt1x} that at density $\rho_0 \sigma^3 = 8.80$ the guessed $R\bar{3}c$
structure has the lowest value of the potential energy between all of the considered structures. 

A characteristic feature of our guessed structure is that not all 
particles in the structure have identical environments.
Thus, the energies of all red and green particles in Fig.\ref{fig:3view-8x8} are equal to each other. 
Energies of all blue particles are also equal to each other. 
However, the energies of the red/green particles are not equal to the energies of the blue particles.

The possibility of formation of the structures in which not all particles have identical local environments, 
while all particles are the same, has been discussed in Ref.\cite{Stillinger20131}.

\subsection{Density $\rho_0 \sigma^3 = 9.60$}

The guessed structure is the BCC lattice with the length
of the edge of the unit cell cube equal to $a = 0.5928$.

We found that the BCC structure is stable on heating 
up to the temperature $T_m \approx 9.875\cdot10^{-3}$.
There were no signs of a transformation of this lattice 
into some other structure in the range of temperatures below $T_m$. 
The comparison of the PDF calculated on the structure obtained 
through crystallization of the liquid with the PDF calculated on the 
BCC lattice at nearly zero temperature is shown in Fig.\ref{fig:pdfs-03}(f)

\section{Energies of the selected lattices as the functions of density}\label{sec:groundstate}

In the previous parts of this paper, we described our observations concerning 
the crystallization of particles interacting through the harmonic-repulsive 
potential into several crystal structures whose formations had not been 
anticipated in Ref.\cite{LuZY20111,Frenkel20091,Prestipino20091}. 
From this perspective, it is of interest to compare how the potential energies 
of the selected optimized crystal structures in the ideal ground 
states depend on density. 
Of course, this comparison is not sufficient to draw conclusions about 
the behaviors of the systems at non-zero temperatures. 
However, such considerations can provide an insight into the behavior 
of the systems at non-zero temperatures \cite{Prestipino20091}. 

\begin{figure}
\begin{center}
\includegraphics[angle=0,width=3.2in]{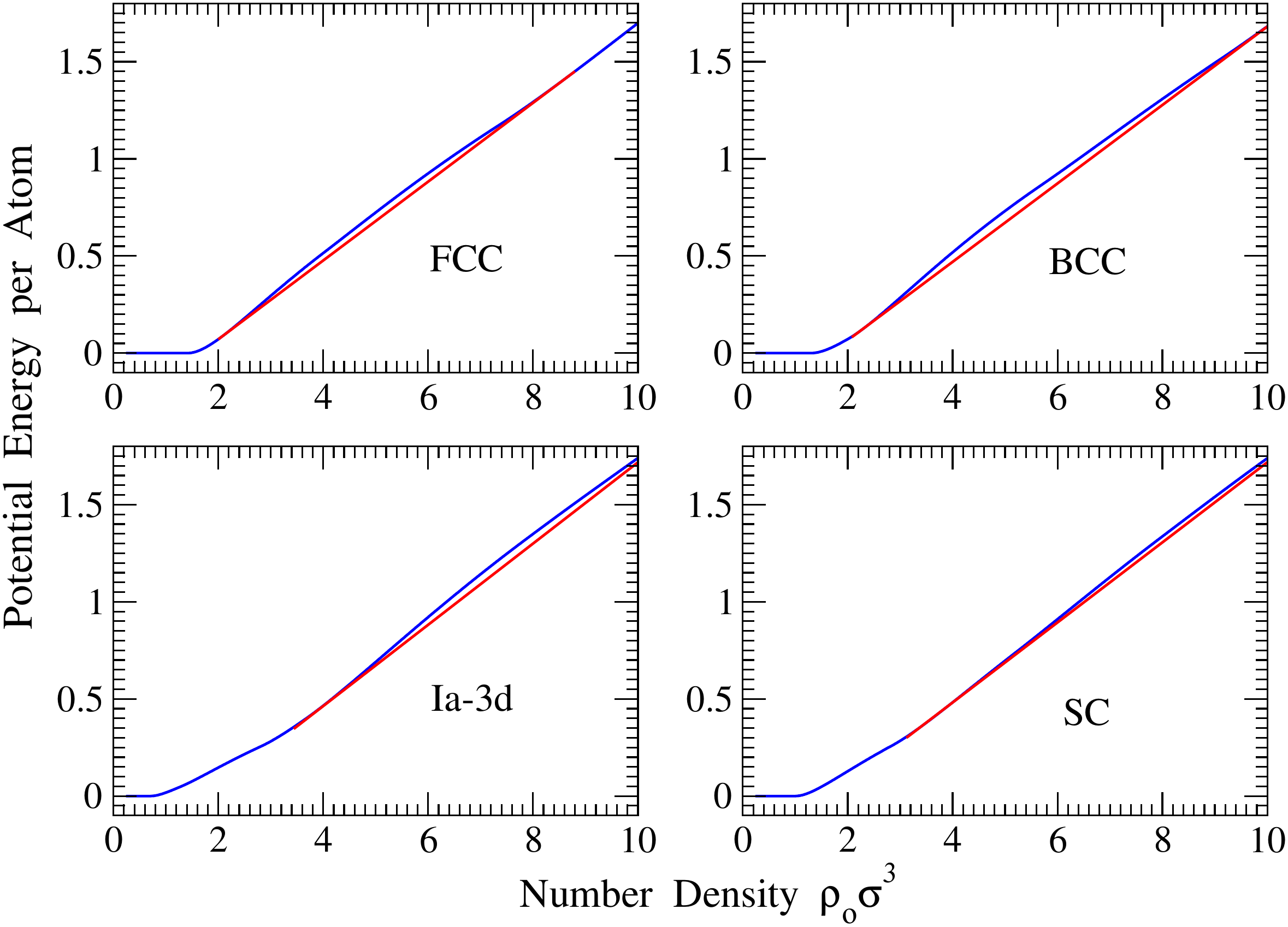}
\caption{
The blue curves show the potential energies of the FCC, BCC, 
$Ia\bar{3}d$, and SC lattices as functions of the density. 
The red straight lines show that the blue curves are concave in certain intervals of density. 
The concave shape of the curves suggests that in certain 
interval of density every considered lattice is unstable 
with respect to the phase separation. 
It is (quite) possible, of course, that some other lattice has even lower value of potential energy
than the particular chosen lattice and that this another lattice is stable with respect 
to the phase separation.
}\label{fig:potlatt1y}
\end{center}
\end{figure}

Since we consider the purely repulsive potential that monotonically increases as 
the distance decreases it is clear that the potential energy of any chosen 
lattice should monotonically increase as the density of the lattice increases. 
This behavior is illustrated in Fig.\ref{fig:potlatt1y}.

Note in Fig.\ref{fig:potlatt1y} that the dependencies of the potential 
energies on the density for the selected lattice are concave 
functions in certain ranges of the density.
Thus, if no other lattices, except one, were possible (hypothetically) 
then it would be favorable for this (the only possible) lattice, to split
into two phases with different densities 
(the two phases are formed by the same crystal lattice, 
but the densities of the two lattices are different). 
However, since other lattices are possible, it may turn out that
the phase separation for any given lattice never occurs because 
another crystal lattice intervenes and precludes this phase separation.

%%%%%%%%%%%%%%%%%%%%%%%%%%%%%%%%%%%%%%
% Begin Figure
%%%%%%%%%%%%%%%%%%%%%%%%%%%%%%%%%%%%%%
%%%%%%%%%%%%%%%%%%%%%%%%%%%%%%%%%%%%%%
% End Figure
%%%%%%%%%%%%%%%%%%%%%%%%%%%%%%%%%%%%%   
%%%%%%%%%%%%%%%%%%%%%%%%%%%%%%%%%%%%%%
% Begin Figure
%%%%%%%%%%%%%%%%%%%%%%%%%%%%%%%%%%%%%%
\begin{figure*}
\begin{center}
\includegraphics[angle=0,width=7.0in]{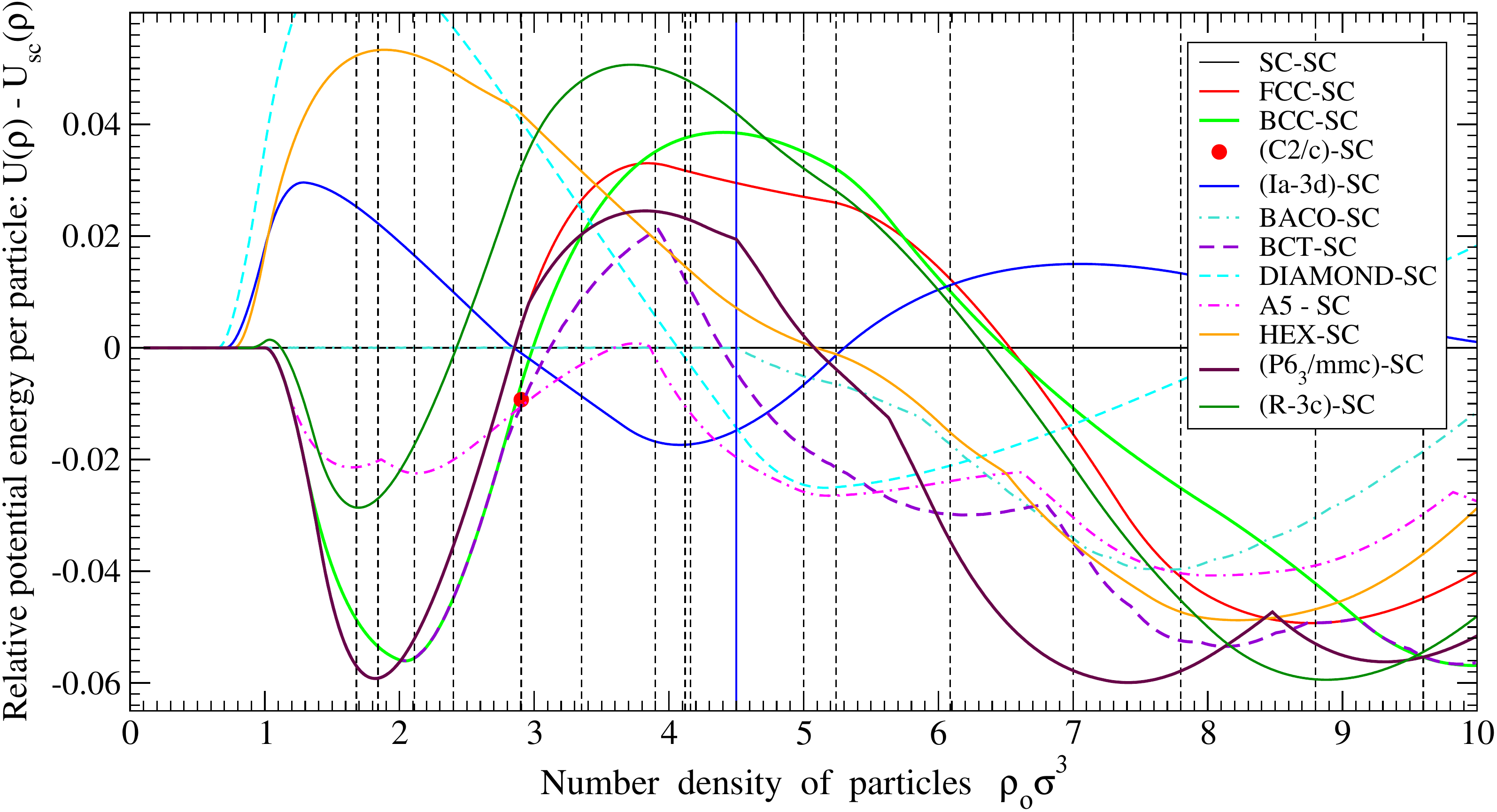}
\caption{
The dependencies of the Potential Energies Per Particle (PEpP) of the selected lattices on the particles' 
number density relative to the PEpP of the Simple Cubic (SC) lattice at the same density. 
The positions of the vertical dashed lines correspond to the densities at which the MD simulations were performed.
The notations for the curves are the following: ``FCC-SC" is the PEpP of the Face Centered Cubic 
lattice minus the PEpP of the Simple Cubic lattice.
The notation $C2/c$ stands for the monoclinic unit cell with 32 particles in the unit cell that generates 
the columnar structure at the density $\rho_o \sigma^3 = 2.904$ (single red filled circle).
The notation $Ia\text{-}3d$ stands for the $Ia\bar{3}d$  
cubic lattice (space group $\#230$) with 16 particles per unit cell occupying the (16b) Wyckoff special positions. 
``HEX"-stands for the Hexagonal lattice. 
``BACO" stands for the Base Centered Orthorhombic lattice. 
``BCT"-stands for the Body Centered Tetragonal lattice. 
It follows from the figure that in the region of densities
$0 \lesssim \rho_o \sigma^3 \lesssim 2.92$ $FCC$ or $BCC$ lattices have lower values of the PEpP than the other lattices.
In the region $2.96 \lesssim \rho_o \sigma^3 \lesssim 4.50$ the $Ia\bar{3}d$ lattice has the lower value 
of the PEpP than the other lattices.
At the values of the density $\rho_o \sigma^3 = 5.0$ and $\rho_o \sigma^3 = 5.24$ our best-guess structure 
is the tetragonal lattice $I4_1/amd$ (space group $\#141$) with four particles per unit cell 
occupying the (4b) Wyckoff special positions. 
This $I4_1/amd$ lattice is also the $A5$ or $\beta Sn$ lattice. 
It is possible to think about this lattice as about the distorted diamond structure. 
The curve corresponding to this lattice was optimized with respect to the ratio $c/a$.
Note that the curve corresponding to the $I4_1/amd$ lattice lies below the curve for the undistorted diamond structure.
In the region $6.10 \lesssim \rho_o \sigma^3 \lesssim 8.10$ the hexagonal lattice $P6_3/mmc$ (space group $\#194$) 
has the lowest value of the PEpP between the considered lattices.
This lattice can be described as formed by two parallel interchanging triangular lattices displaced 
with respect to each other in the planes of the triangular lattices, as in the hexagonal lattice. 
The separation between the two nearest triangular planes in the guessed and optimized $P6_3/mmc$ lattice is, 
however, different from the separation that occurs in the hexagonal close-packed lattice. 
At the density $\rho_o\sigma^3 = 8.8$ the lowest value of the PEpP has the 
guessed and optimized hexagonal lattice $R\bar{3}c$ (space group $\#167$) 
with 24 particles per unit cell occupying (6a) and (18e) Wyckoff special positions. 
}\label{fig:potlatt1x}
\end{center}
\end{figure*}
%%%%%%%%%%%%%%%%%%%%%%%%%%%%%%%%%%%%%%
% End Figure
%%%%%%%%%%%%%%%%%%%%%%%%%%%%%%%%%%%%% 

Figure \ref{fig:potlatt1x} shows how the PEpP 
of the selected crystal lattices differ from the PEpP of the simple cubic lattice at the same value of density.
At any value of the density the lattice with the lowest value of the PEpP should be the
most stable between the considered lattices in the NVT ensemble at zero temperature. 

It is of interest that at any chosen value of density from table \ref{table:MD-summary-1} 
the curve that has the lowest value of the potential energy at this density corresponds 
to the lattice that we guessed from the analysis of crystal structures obtained from the MD simulations. 
This situation shows that our guesses of the crystal structures indeed can correspond 
to the ground states at the corresponding densities. 
Additionally, this shows that the zero-temperature considerations indeed can shed some 
light on the behavior of systems at non-zero temperatures, 
as it has been assumed also in Ref.\cite{Prestipino20091}

\section{Selected results from the NPT simulations}\label{sec:NPT-simulations}

In order to further address the stabilities of the observed crystal lattices, 
we performed constant pressure (NPT) simulations starting from the crystal structures 
obtained in the NVT simulations. 
Every NPT simulation has been carried out at a constant value of temperature. 
We slowly varied the pressure within the LAMMPS program with the damping parameter 
set to $1000$ MD steps.
We found that the structures that we obtained in the NVT simulations remained stable in 
the NPT simulations in certain ranges of pressure. 
We monitored the stability of the lattices through the dependencies of the PEpP and the density 
on pressure. 
At the borders of the stability regions the mentioned dependencies exhibit discontinuities in their slopes. 
Of course, the results obtained in these NPT simulations do not establish the true regions 
of stabilities for the discussed lattices. 
However, they do provide a certain insight into the regions of the lattice stabilities.  
The results of these simulations are summarized in Table \ref{table:NPT-summary-1}.

\begin{figure}
\begin{center}
\includegraphics[angle=0,width=3.2in]{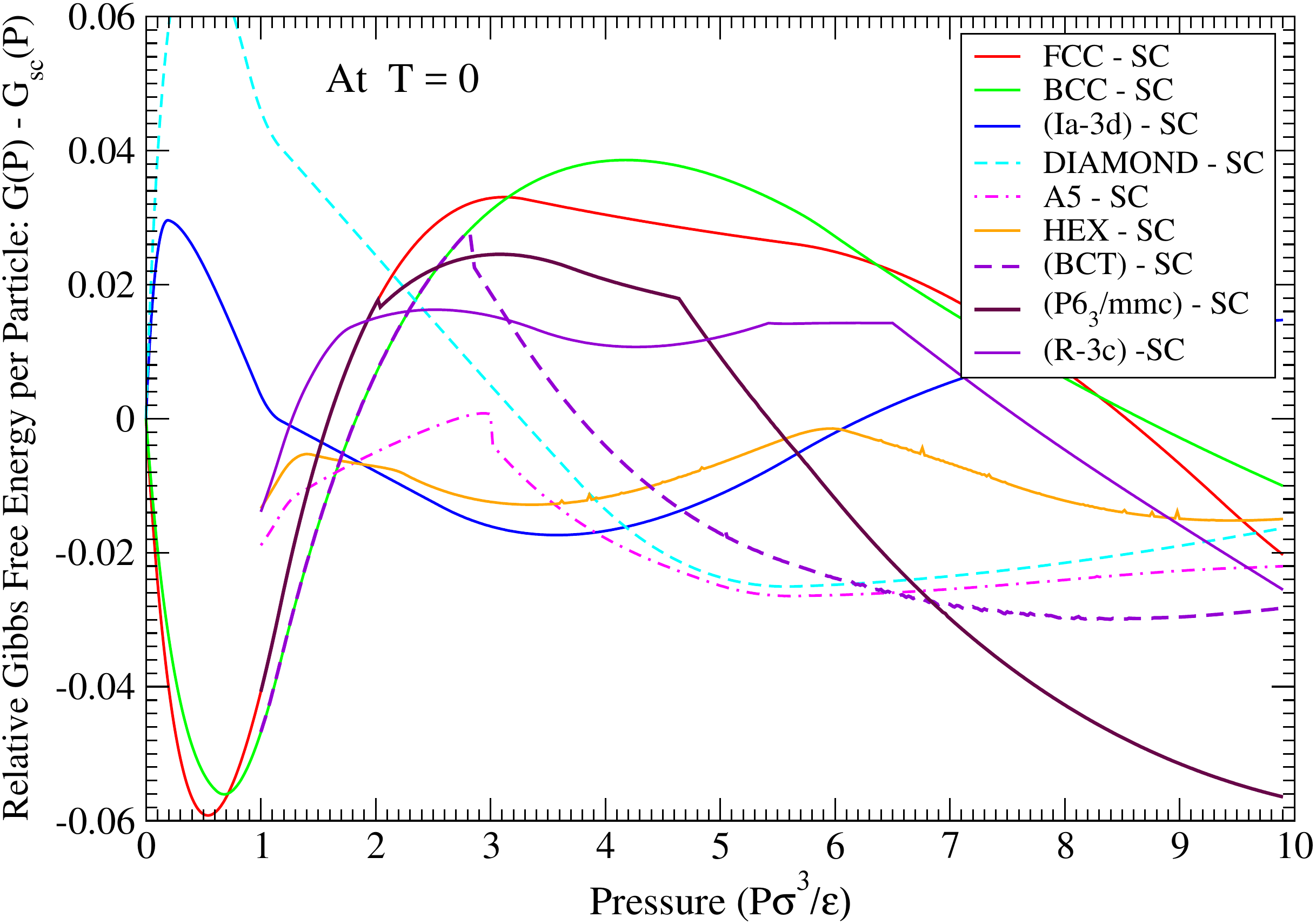}
\caption{
The dependencies on pressure at $T=0$ of the differences between the Gibbs 
free energies (chemical potentials) for the selected lattices and the Gibbs 
free energy for the simple cubic lattice.
Note that at very low pressure the FCC lattice has the lowest value of the Gibbs free energy, 
in accordance with the results presented in Table \ref{table:MD-summary-1}. 
As pressure increases the BCC lattice becomes more stable than the FCC lattice. 
As pressure increases further the $Ia\bar{3}d$ becomes the most stable 
between the considered lattices.
Note, however, that we did not calculate the Gibbs free energy for 
the $C2/c$ structure whose region 
of stability can be expected to occur between the regions of stability 
for the BCC and the $Ia\bar{3}d$ lattices. On further increase of pressure 
the $A5$ (i.e., the distorted diamond) structure becomes more stable than the $Ia\bar{3}d$ crystal structure, 
in agreement with Table \ref{table:MD-summary-1}. Then the $P6_3/mmc$ lattice becomes the most stable.
}\label{fig:gibbs}
\end{center}
\end{figure}

The Gibbs free energy, $\Phi = U - TS + PV$, should be at the global 
minimum in the equilibrium simulations at constant pressure
($S$ is the entropy of the system). At zero temperature we have $\Phi = U + PV$.  
To address how reasonable are the results presented in Table \ref{table:NPT-summary-1}, 
we calculated for the selected lattices how their Gibbs free energies depend on 
pressure at zero temperature. While the results of such calculations are strictly 
applicable only at zero temperatures, they nevertheless provide important intuitive 
insights into the phase diagrams at non-zero temperatures \cite{Prestipino20091}.
In order to calculate the Gibbs free energy at zero temperature we varied the parameters 
of the lattices, choosing those that have the required value of pressure, 
and then selected from those the lattice parameters that lead to the lowest value of the Gibbs free energy.
The results of these calculations are shown in Fig.\ref{fig:gibbs}.
We see in Fig.\ref{fig:gibbs} that the curves with the lowest values of 
the Gibbs free energy in some intervals of pressure almost always correspond to the lattices which were observed as stable, 
according to Table \ref{table:NPT-summary-1}, in the NPT simulations.

\begin{center}
\begin{table}
\begin{tabular}{| c | c | c | c | c | c | c | c | c |} \hline
$\rho_o \sigma^3$       & $1.75$& $2.20$ & $2.904$ & $3.60$&$5.12$& $6.10$    & $8.8$  \\\hline
Lattice                 & $FCC$ & $BCC$  & $C2/c$  & $Ia\bar{3}d$& $A5$& $2H$& $R\bar{3}c$  \\\hline
$(T/\epsilon)10^3$& $5.0$ & $6.0$  & $3.0$   & $4.0$& $3.2$& $3.50$  & $4.50$  \\\hline
$P_L(\sigma^3/\epsilon)$&$0.15$&$0.55$& $1.60$&$1.75$& $4.50$& $5.55$& $12.75$  \\ \hline
$P_H(\sigma^3/\epsilon)$&$0.62$&$1.67$& $1.92$&$4.45$& $6.00$& $8.65$& $21.50$  \\ \hline
\end{tabular}
\caption{
The 1st row in the table shows the densities at which the starting crystal 
structures have been produced in the NVT simulations from the liquid state. 
The 2nd row shows the types of the starting crystal structures. 
The notation $A5$ is used for the $I4_1/amd$ lattice, while 
the notation $2H$ is used for the $P6_3/mmc$ lattice formed 
by 2 alternating Hexagonal lattices. 
The 3rd row shows the temperatures at which the NPT simulations have been performed.
The 4th and the 5th rows show the lowest ($L$) and the highest ($H$) approximate 
values of the pressure at which the lattices remained stable in the NPT simulations.
}
\label{table:NPT-summary-1}
\end{table}  
\end{center}

%%%%%%%%%%%%%%%%%%%%%%%%%%%%%%%%%%%%%%
% Begin Figure
%%%%%%%%%%%%%%%%%%%%%%%%%%%%%%%%%%%%%%
%%%%%%%%%%%%%%%%%%%%%%%%%%%%%%%%%%%%%%
% End Figure
%%%%%%%%%%%%%%%%%%%%%%%%%%%%%%%%%%%%%   

\section{Conclusions}\label{sec:conclusion}

We investigated the behavior of particles interacting through the harmonic-repulsive 
pair potential at different number densities using direct MD simulations. 
At several densities we observed behaviors that have not been anticipated previously. 
The results of a particular interest are the following:

1) At the density $\rho_o \sigma^3 = 2.904$ we observed significant resilience of the liquid against
crystallization. Yet, we have been able to observe crystallization
from the liquid state into a $C2/c$ monoclinic structure with 32 particles in the
unit cell occupying four different Wyckoff (8f) sites.
We found that particles at the different Wyckoff sites have different values of the potential energy. 
The possibility of formation of structures with identical particles occupying
positions with different local environments has been discussed in Ref.\cite{Stillinger20131}

It is possible to think that particles in the observed structure are organized into columns 
where each column is formed by 7 (seven) linear chains of individual particles.
Alternatively, one can also think that each column is formed by 3 (three) helical coils such that 
the full pitch of every coil involves 7 (seven) particles.
From this perspective, the observed structure resembles locally the organization of particles
in some columnar quasicrystals \cite{Dzugutov19931,Engel20071,Ryltsev20151,Ryltsev20171,Damascento20171}.

In our view, further investigations of the liquid state and crystalline structures around 
the density $\rho_o \sigma^3 = 2.904$ are of interest.

2) At density $\rho_o \sigma^3 = 3.352$ we observed crystallization from the liquid 
state into the cubic $Ia\bar{3}d$ (space group $\#230$) crystal structure with 16 particles per unit cell 
occupying the (16b) Wyckoff special positions. This crystal structure has not been 
observed previously in experiments or in computer simulations of single atomic 
or single component systems of particles interacting through pair potentials. 
However, the $Ia\bar{3}d$ crystal structures were observed in 
more complex systems \cite{Luzzati1968,Kutsumizu2012,IsraelachviliJ2011,Tschierske2013,Seddon2006,Cho20151}.

3) At the density $\rho_o \sigma^3 = 4.400$ we were not able to observe crystallization 
despite careful investigations at different temperatures in rather long simulation runs. 
This result is of interest because this behavior was observed in the single component 
system of particles (usually single component systems easily crystallize). 
However, recently there were reports about the absence of crystallization 
in single component systems of particles interacting through more complex potentials 
than the harmonic-repulsive potential used in the current study \cite{Ryltsev20131,Damascento20171}.

4) Our analysis of the $R\bar{3}c$ structure formed at high density suggests 
that we (again) observed a structure in which not all particles have equivalent atomic environments. 
The possibility of formation of such structures has been discussed in Ref.\cite{Stillinger20131}.

At a number of densities our results appear to be in disagreement with 
the previously predicted phase diagram for the harmonic-repulsive potential \cite{LuZY20111}.
In our view, the reason for the disagreement is that the investigations 
in Ref.\cite{LuZY20111} were based on the considerations of a certain set of possible 
crystal structures (a relatively wide and reasonable set). 
However, several structures that we observed in our simulations were not included in this set. 
Thus, in our view, nature essentially outwitted the initial guess of the possible crystal structures.
This point of view is supported by the considerations presented in Ref.\cite{Prestipino20091}.
Further investigations clarifying the origin of disagreements with Ref.\cite{LuZY20111}
and establishing the regions of stabilities of different structures are necessary.

The results presented in this paper open several obvious routes for further investigations. 
These include more detailed investigations of the structural and dynamical properties at the 
selected values of density or pressures for the harmonic-repulsive and similar potentials.
Further investigations of the phase diagrams of the harmonic-repulsive and other similar potentials
are also of interest.\\

\section{Acknowledgements}

The crystallographic classification of the guessed crystal structure presented in 
Fig.\ref{fig:shell-units-organization} 
has been done by M.V. Lobanov and K.A. Lokshin.
We would like express to them our gratitude for their help.

A more detailed crystallographic description of the $Ia\bar{3}d$ 
crystal structure with particles occupying the (16b) Wyckoff 
positions has been presented in a separate publication \cite{Lobanov-Lokshin}.

We also would like to thank K.A. Lokshin, N.V. Podberezskaya,
and M.V. Lobanov and for the useful discussions.

Computer simulations for this work have been partly performed 
on the computer cluster of Novosibirsk State University.

%%%%%%%%%%%%%%%%%%%%%%%%%%%%%%%%%%%%%%%%%%%%%%%%%%%
%%%%%%%%%%%%%%%%%%%%%%%%%%%%%%%%%%%%%%%%%%%%%%%%%%%
%           References     
%%%%%%%%%%%%%%%%%%%%%%%%%%%%%%%%%%%%%%%%%%%%%%%%%%%                    
%%%%%%%%%%%%%%%%%%%%%%%%%%%%%%%%%%%%%%%%%%%%%%%%%%%
%%%%%%%%%%%%%%%%%%%%%%%%%%%%%%%%%%%%%%%%%%%%%%%%%%%
%%%%%%%%%%%%%%%%%%%%%%%%%%%%%%%%%%%%%%%%%%%%%%%%%%%


\begin{thebibliography}{10}
\bibitem{Levashov20162c}
Our goal was to investigate correlations between the eigenvalues of the atomic level stress matrices.
In particular, the similarity in the probability distributions 
of $\lambda_2/\lambda_1$ and $\lambda_3/\lambda_2$,
where $\lambda_1,\lambda_2,\lambda_3$ are 
the eigenvalues of the atomic level stress matrices.
\bibitem{heat-cryst-1}
Sometimes in order to produce a better quality crystalline state we also
heated the crystal formed from the liquid above the observed ``crystallization" 
temperature. Usually we were able to heat the crystal to the temperatures 
significantly above the observed ``crystallization" temperature. 
This trick, however, combined with the consequent cooling to near zero 
temperature did not usually lead to the crystal structures
which were significantly better than the crystal structures obtained without 
after-crystallization heating.
\end{thebibliography}
\end{document}